\documentclass[3p]{elsarticle} 
\makeatletter
\def\ps@pprintTitle{%
 \let\@oddhead\@empty
 \let\@evenhead\@empty
 \def\@oddfoot{\centerline{\thepage}}%
 \let\@evenfoot\@oddfoot}
\makeatother

\date{}
\def\texpsfig#1#2#3{\vbox{\kern #3\hbox{\includegraphics{#1}\kern #2}}\typeout{(#1)}}

\usepackage{url,hyperref}
\usepackage{bbm}
\usepackage{eurosym}                           
\usepackage[latin1]{inputenc}                  
\usepackage{url}                               
\usepackage{longtable}                         
\usepackage{array}                             
\usepackage{graphicx,color}
\usepackage{amsthm}
\usepackage{amsbsy}

\usepackage{amssymb}
\usepackage{amsmath}
\usepackage{enumerate}
\usepackage{graphicx,color}
\usepackage{epsfig}
\usepackage{multirow,bigdelim}
\usepackage[table]{xcolor}
\usepackage{natbib}
\usepackage{booktabs}

\theoremstyle{plain}
\newtheorem{thm}{Theorem}[section]

\newtheorem{definition}[thm]{Definition}
\newtheorem{proposition}[thm]{Proposition}

\newtheorem*{rem}{Remark}
\theoremstyle{remark}

\theoremstyle{plain}

\theoremstyle{definition}

\newcommand{\e}{{\rm e}}        
\def\E{\mathbb{ E}}             
\def\Q{\mathbb{ Q}}             

\renewcommand{\d}{{\rm d}}      
\def\dW{{\rm d}W}               
\def\dt{{\rm d}t}

\newcommand{\CPR}{\Lambda}
\newcommand{\SMM}{\text{SMM}}

\newcommand{\RI}{\text{RI}}

\def\1{{\mathbbm{1}}}            

\theoremstyle{plain}

\usepackage[margin=1cm]{caption}

\DeclareMathOperator*{\argmin}{arg\,min}

\usepackage{algorithm}
\usepackage[noend]{algpseudocode}

\usepackage{tikz}
\usetikzlibrary{arrows,chains,matrix,positioning,scopes}
\makeatletter
\tikzset{join/.code=\tikzset{after node path={%
\ifx\tikzchainprevious\pgfutil@empty\else(\tikzchainprevious)%
edge[every join]#1(\tikzchaincurrent)\fi}}}
\makeatother
\tikzset{>=stealth',every on chain/.append style={join},
         every join/.style={->}}
\tikzstyle{labeled}=[execute at begin node=$\scriptstyle,
   execute at end node=$]

\numberwithin{equation}{section}	     

\title{Pricing and Hedging Prepayment Risk in a Mortgage Portfolio}

\begin{document}

\author[1]{Emanuele Casamassima}
\ead{emanuele@block-tech.io}
\author[1,3]{Lech A.~Grzelak\corref{cor1}}
\ead{L.A.Grzelak@uu.nl}
\author[1]{Frank A. Mulder}
\ead{Frank.Mulder@rabobank.com}
\author[3]{Cornelis W.~Oosterlee}
\ead{C.W.Oosterlee@uu.nl}
\cortext[cor1]{Corresponding author at Rabobank, Utrecht, the Netherlands.}
\address[1]{Rabobank, Utrecht, the Netherlands}
\address[3]{Utrecht University, Utrecht, the Netherlands}

\begin{abstract}
    \noindent
Understanding mortgage prepayment is crucial for any financial institution providing mortgages, and it is important for  hedging the risk resulting from such unexpected cash flows. Here, in the setting of a Dutch mortgage provider, we propose to include non-linear financial instruments in the hedge portfolio when dealing with mortgages with the option to prepay part of the notional early. Based on the assumption that there is a correlation between  prepayment and the interest rates in the market, a model is proposed which is based on a specific refinancing incentive. The linear and non-linear risks are addressed by a set of tradeable instruments in a static hedge strategy. We will show that a stochastic model for the notional of a mortgage unveils non-linear risk embedded in a prepayment option. Based on a calibration of the refinancing incentive on a data set of more than thirty million observations, a functional form of the prepayments is defined, which accurately reflects the borrowers' behaviour. We compare this functional form with a  fully rational model, where the option to prepay is assumed to be exercised rationally.
\end{abstract}

\maketitle


\section{Introduction}

{\let\thefootnote\relax\footnotetext{The views expressed in this paper are the personal views of the authors and do not necessarily reflect the views or policies of their current or past employers.}}

A mortgage is a long-term loan that is secured by a registered good. The two counterparties of a mortgage are the lender (the \textit{mortgagee}), and the borrower (\textit{mortgagor}).
The mortgages that we consider here (that are common products in the Netherlands) provide a mortgagor with two embedded options: regarding the choice of the mortgage's interest rate and the possibility to deviate from the scheduled, future cash flows. The first option gives rise to so-called  \textit{pipeline risk}: the borrower has the opportunity to get a mortgage based on the lowest rate from the grace period, which usually is three months. The second option generates the {\it prepayment risk}, and is our current interest.

In this paper, we will present a new hedging strategy for a financial institution to deal with the prepayment risks, i.e., to prepay the notional and major parts of the remaining interest rates.
When a mortgage is settled, the borrower obtains a precise schedule of payments that has to be followed. These cash flows guarantee that the borrower ultimately pays back the initial sum, i.e., the notional, plus an extra amount of money representing the cost of the loan. These payments are called {\it repayments}. Mortgages come with an embedded option that gives the possibility to the mortgagor to {\it prepay}, i.e., to redeem part of the debt in advance, thus deviating from the scheduled plan of amortization.

Prepayment risk can be substantial, as a financial institution typically relies on long-term payment periods by the mortgagor, with corresponding interest rate payments. 
When a mortgagor signs a mortgage contract with a bank, the bank typically sets up a deal with another financial institution to collect the lump sum for immediate payment to the client, and mirrors the cash flows that will occur with the mortgagor. When the contract details change due to prepayment, this may have a significant impact on the cash-flows in the context of this mortgage. The interest, which depends on the interest rate, notional and duration of the loan, compensates for the bank's costs and partly represents the bank's profit from selling the mortgage. Prepayment could result in a loss, because the notional and the interest rate may get lower, and the loan duration may be reduced. When a large number of bank clients suddenly prepay, significant prepayment risk may result for the bank, which needs to be analyzed and hedged. In this paper, we focus particularly on the hedging of the prepayment option. We propose a static hedging strategy here, in which linear and a nonlinear financial products are used.
We will show that the risks associated with prepayments can be mitigated by means of volatility-sensitive interest rate products (assuming that the refinancing incentive is mainly driven by interest rates observed in the financial market). 


The paper is organized as follows. After discussing some basic notions, like some well-known mortgage contracts, and, particularly, the refinancing incentive, in Section~\ref{2}, we will present our pricing model in Section~\ref{3}. There, a link between the value of the mortgage portfolio and the level of the interest rates in the market is established. The Index Amortizing Swap (IAS) will play a prominent role in this. Moreover, in Subsection~\ref{da}, the intensity of prepayments, which is known as the {\it Conditional Prepayment Rate} (CPR), will be defined.
In Section~\ref{4}, we then  perform computations with the IAS based on a deterministic CPR function. Hedge strategies, based on a stochastic conditional prepayment rate, are discussed and analyzed in detail in Section~\ref{5}, based on several numerical experiments. Section~\ref{6} concludes.


\section{Preliminaries}
\label{2}
In this section, we provide some details about typical mortgage structures and information about prepayment incentives.

Mortgages are classified according to the amortization plan that the notional follows during the lifetime of the contract. With all other characteristics equal (duration, initial notional, interest rate), the amortization plan influences the total amount of interest raised  and how the notional is paid back. Different amortization plans also give rise to a different impact on the prepayment rate. Two of the most commonly sold mortgage plans are the {\it bullet mortgage} and the {\it annuity mortgage}.

\subsection{Bullet}
\label{sec:bullet}
The bullet is a straightforward mortgage, where the mortgagor receives $N_0$ at the time of settling, $t_0$, and the notional is fully redeemed at the end of the contract period, in one single payment. At the end of each payment period, only the interest part is paid to the mortgagee, so the notional remains constant until the final time $T_M$, i.e.,
$
    N(T_i) = N_0 \mathbbm{1}_{\{ T_i < T_M \}}.
$
The installment, $C(T_i)$, serves to pay the interest part that is due at time $T_i$, which is based on the notional of the loan, $N_0$, the interest rate, $K$, and the time span that the payment covers, i.e., $    C(T_i) = K N_0 \tau_i,$ with $\tau_i=T_{i+1}-T_i.$
The total amount of interest that a lender receives at the end of the contract equals $I = \sum_{i=1}^{M} K N_0 \tau_i.$ We will use $\Lambda(t)$ to denote the Conditional Prepayment Rate (CPR). If we assume a CPR within the time span to be constant, i.e.,  $\Lambda(t)\equiv \Lambda$, and note that in a bullet the repayments are equal to zero, the notional at time $T_i$ is given by:
\begin{equation*}
    N(T_i) = (1 - \Lambda) N(T_{i-1}) =(1 - \Lambda)^2  N(T_{i-2}) = ... = (1 - \Lambda)^i  N_0,
\end{equation*}
with the total amount of interest,
\begin{equation*}
    I = \sum_{i=0}^{M-1} K N(T_i) = K N_0 \sum_{i=0}^{M-1} (1 - \Lambda)^i = \frac{K N_0}{\Lambda} (1 - (1 - \Lambda)^M).
\end{equation*}

\begin{figure}[h]
    \centering
    \includegraphics[scale=0.43]{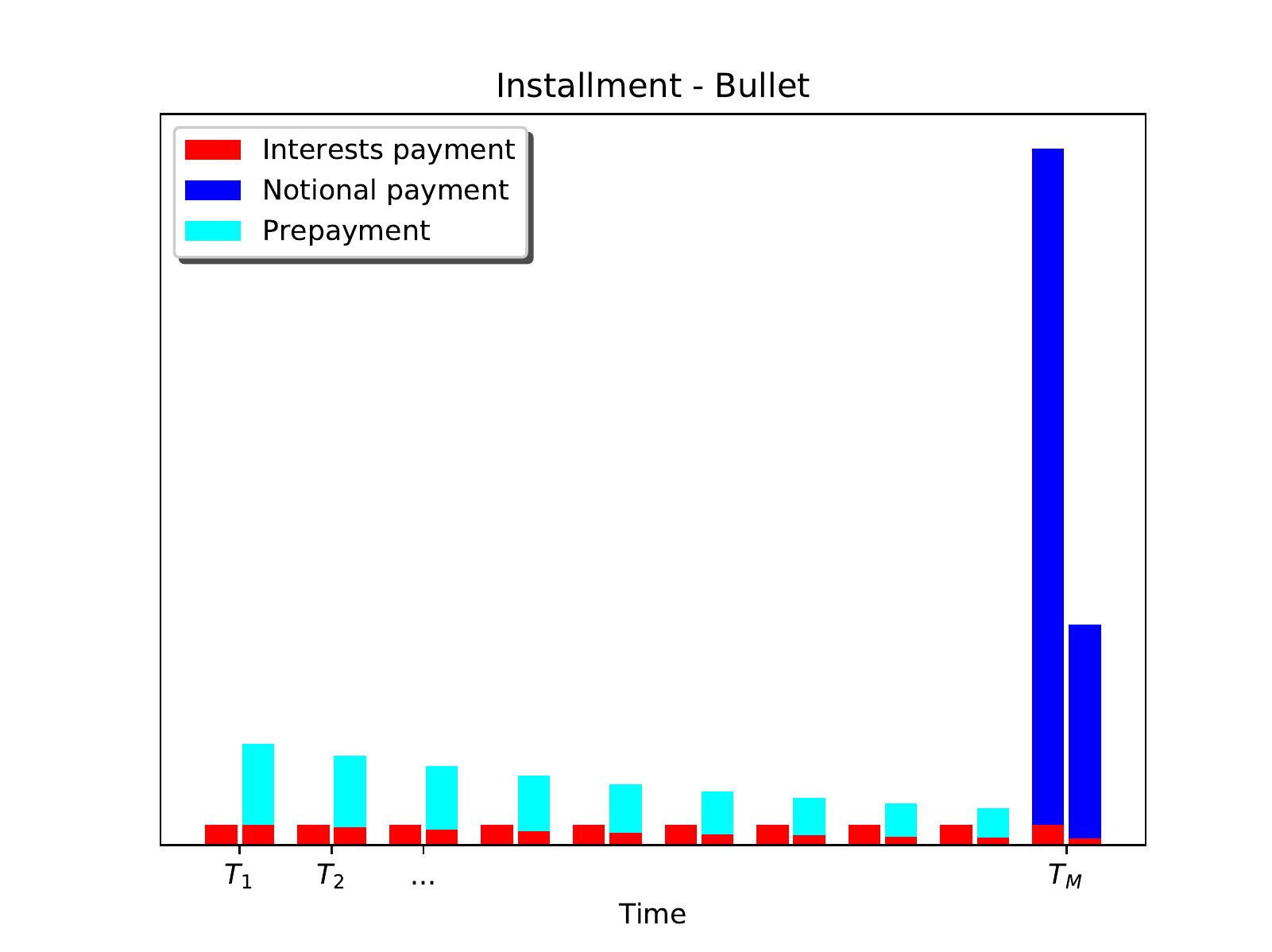}
    \includegraphics[scale=0.43]{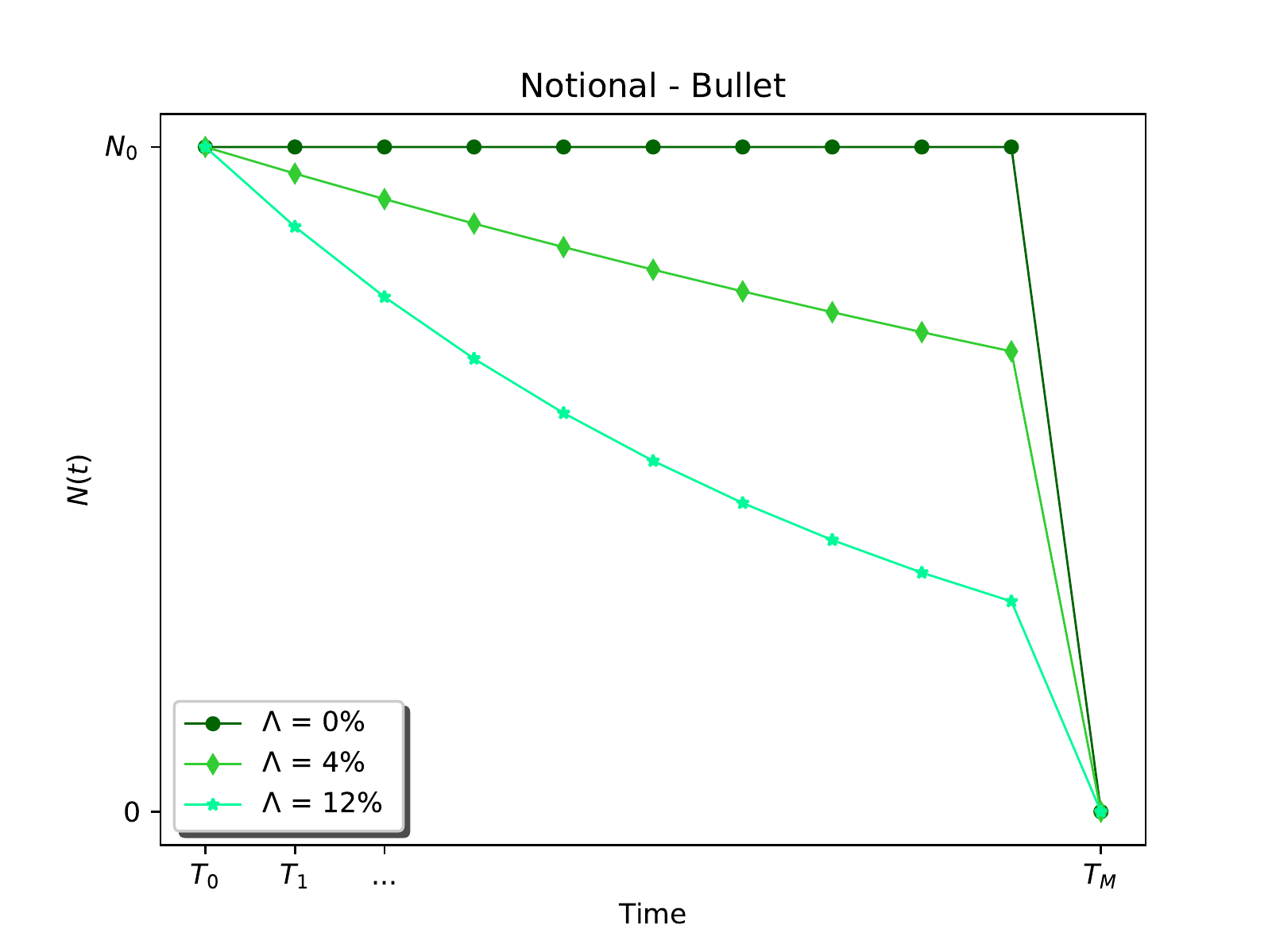}
    \caption{Bullet mortgage payment profile with $T_M = 10$ and $K = 3\%$ under different scenarios. Left: the installment composition in case $\Lambda = 0\%$ or $\Lambda = 12\%$. Right: outstanding notional in time under different levels of prepayment.}
    \label{fig:Bullet_DifferentScenarios}
\end{figure}
In Figure \ref{fig:Bullet_DifferentScenarios}, we see how the prepayments would impact the bullet mortgage. The maturity in this example is set to 10 years, and the mortgage rate is $3\%$. In the left-side figure, the installment that the bank receives is presented for two cases, one without prepayments and another with $\Lambda = 12\%$. In the first case, the installment is constant, as this contract does not involve any repayment. In the other case, we see how the prepayments (represented by the cyan bars) are added to the contractual payments of the interest.

As a consequence, the notional diminishes in time, and so do the interest payments based on it. The final installment is therefore not as large as in the case of no prepayments. In the right-side graph, the effect of $\Lambda$ on the remaining notional $N(t)$ is shown. Clearly, the higher the prepayment rate, the stronger the decay of the notional in time.

\subsection{Annuity}
\label{sec:annuity}
An annuity is a somewhat more involved contract because, contrary to the bullet, it involves repayments. The repayment $Q(T_i)$ diminishes the notional by the same amount, i.e.,
\begin{equation}
    N(T_{i+1}) = N(T_i) - \Delta T_i Q(T_i) = N(T_i) - \Delta T_i \big( C(T_i) - I(T_i) \big).
    \label{eqn:NotionalUpdateAnnuity}
\end{equation}
When a first payment has taken place, one year after signing the contract, the annuity is called an \textit{ordinary annuity}, whereas, if the first amount was paid immediately, it would be an \textit{annuity due}, see also \cite{fabozzi2005fixed}. Our focus is on the ordinary annuity, which we will simply  call annuity.

An essential characteristic of an annuity is that the installments, $C(T_i)$, are fixed amounts, i.e., $C(T_i) \equiv C$,
with $i=1,...,M$ (representing an equidistant partitioning of the time interval in years). The interest rate and the principal parts have to be balanced so that the sum is constant at each payment date. Therefore, these quantities will follow opposite trends. When the notional is progressively paid back, the interests computed on the notional will get smaller.
To calculate the correct installment amount $C,$ we impose that the present value of all future installments should be equal to the notional of the mortgage, so with time intervals of one year, we can write the annuity as,
\begin{eqnarray}
    \text{An}(t_0;K) = \sum_{i=1}^M \frac{C}{(1 + K)^{T_i}} &=& \frac{C}{(1 + K)} \sum_{i=0}^{M-1} \frac{1}{(1 + K)^{T_i}} \nonumber \\
    &=& \frac{C}{K} \left(1 - \frac{1}{(1 + K)^{T_M}} \right) \equiv N_0,
    \label{eqn:PresentValueAnnuity}
\end{eqnarray}
and thus,
\begin{equation}
    C = \frac{K N_0}{1 - (1 + K)^{-{T_M}}}.
    \label{eqn:InstallmentAnnuity}
\end{equation}
With this, we derive the interest rate payment, $I(T_i)=K N(T_{i-1})$, and the principal payment, $Q(T_i) =  C(T_i) - I(T_i)$.

\begin{figure}[h]
    \centering
    \includegraphics[scale=0.43]{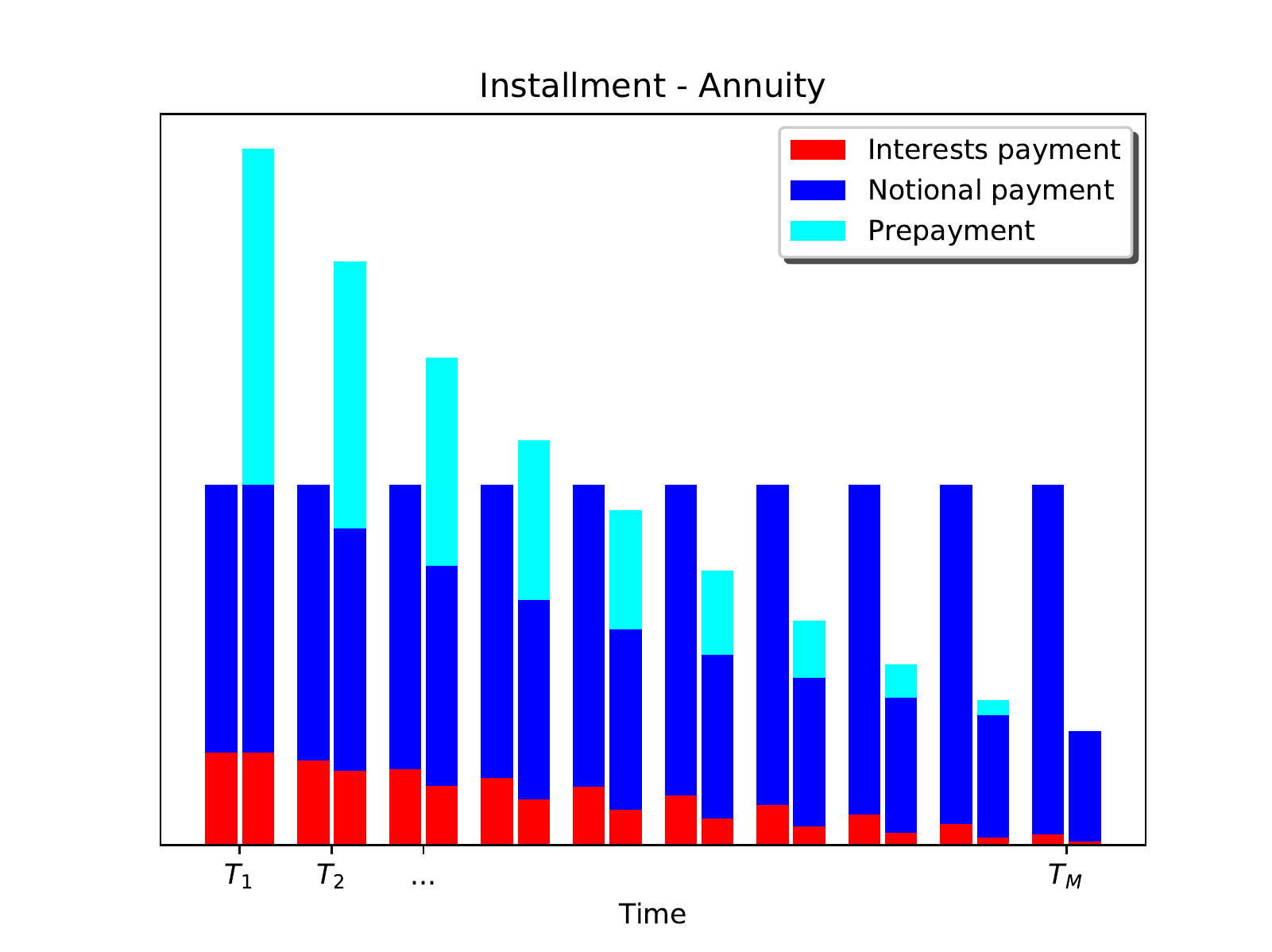}
    \includegraphics[scale=0.43]{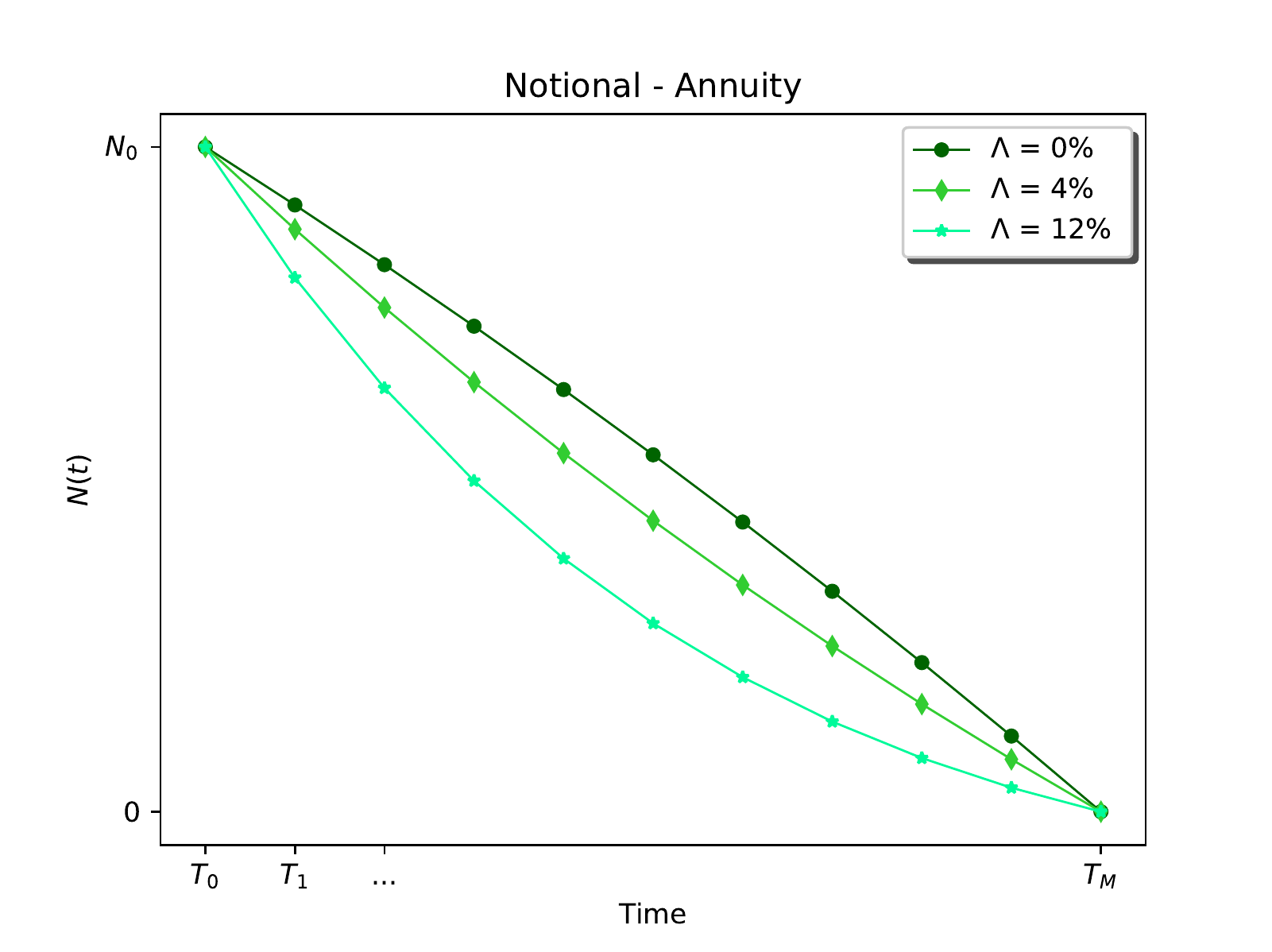}
    \caption{Annuity with $T_M = 10$ and $K = 3\%$ under different scenarios. Left: the installment composition in case $\Lambda = 0\%$ or $\Lambda = 12\%$. Right: the outstanding notional in time under different levels of prepayment.}
    \label{fig:Annuity_DifferentScenarios}
\end{figure}

Including the CPR $\Lambda$, as in the case of the bullet, gives us,
\begin{equation}
    N(T_{i+1}) = N(T_i) - Q(T_i) - \Lambda \left(N(T_i)-Q(T_i)\right).
    \label{eqn:NotionalUpdateAnnuity_ConstantPrepayment}
\end{equation}
The quantity $\Lambda(T_i)$, at time $T_i$, can also be interpreted as a reformulation of the interest payment $I(T_{i+1})$ and the installment $C(T_{i+1})$. When a mortgagor decides to prepay, the installment for the remaining dates is rebalanced according to the updated outstanding notional. Consequently, $C(T_i)$ becomes a time-dependent quantity,
\begin{equation}
    C(T_i) = \frac{K N(T_i)}{1 - (1 + K)^{-(T_M - T_i)}}.
    \label{eqn:InstallmentAnnuity_ConstantPrepayment}
\end{equation}
Examples of annuity payments are presented in Figure \ref{fig:Annuity_DifferentScenarios}. As for the bullet, in the left-side graph, we compare the coupon magnitude for $\Lambda = 0\%$ and $\Lambda = 12\%$, specifying with different colors the different components. In the right-side chart, the impact of varying prepayment levels on the outstanding notional is shown. The larger the $\Lambda$, the greater the rate of reduction.


\subsection{Prepayment determinants}
Regarding the incentives to prepay a mortgage plan, people do not always act rationally, meaning that even if there is a strong incentive for prepayment, mortgagors may not prepay, and vice versa. This fact has an impact on  the type of model that will be chosen to forecast prepayments. There are two main lines of prepayment models: \textit{optimal prepayment models} and \textit{exogenous models}. Optimal prepayment models (that are also called fully rational models here) rely on the assumption that people behave rationally, and will prepay when the value of their mortgage is greater than the outstanding debt (including penalties for refinancing). Examples can be found in \cite{van1998valuation}, \cite{kuijpers2007optimal}, and \cite{sherrispricing}. The problem with these models is that they only depend on the interest rate level in the market, leaving aside any endogenous variables (for instance, the mortgagor's age, the mortgage age, period of the year).

The exogenous models can be subdivided into {\it extended endogenous} and {\it strictly empirical} models. The first  type is based on the same modeling approach as the optimal prepayment models, but prepayments need not be directly caused by the level of the interest rates. Moreover, the approach also involves European options to address the prepayment risk. Examples of this type can found in \cite{dunn1981comparison} and \cite{stanton1995rational}, amongst others.
The purpose of the strictly empirical models is to attribute the prepayments to a set of variables that are assumed to be correlated to prepayment. The authors in \cite{alink2002mortgage} analyse the pros and cons of the two main approaches of this kind being so-called survival analysis and logistic regression approaches.

Mortgage rates are fluctuating quantities that different financial institutions quote, and they depend on the type of mortgage, the contract's maturity, and often on the type of house used as the collateral.
Research on variables that significantly influence the prepayment rate can be found in~\cite{charlier2003prepayment, hoda2007implementation, bandic2004pricing, consalvi2010measuring, van1998valuation, alink2002mortgage, jacobs2005modelling, bissiri2014modeling, richard1989prepayments, kalotay2004option, perry2001study, castagna2013measuring, fabozzi1992mortgage, kolbe2008valuation, Agarwal}. There may be multiple reasons for prepayment that are very different, however, there is agreement about one specific driver, i.e., {\it the refinancing incentive}.
The refinancing incentive occurs when a mortgagor observes a lower rate than the rate on her mortgage.  Other well-known drivers that have a natural explanation are the mortgage age, or the month of the year.
Based on the available data, other variables may be the mortgagor's age, housing turnover, or the amount on one's bank account.

\subsubsection{Functional form for the refinancing incentive}
The interest rate incentive is one of the main reasons to prepay. In this section, we will define a suitable model for this.

A reasonable benchmark for the price of a mortgage may be the swap rate, $S_{t,T}(t)$, which matches the maturity and the frequency of payments of the mortgage, where a spread is added. Banks typically derive the at-the-money mortgage rate for new clients from the present level of the corresponding swap rates. The initial mortgage rate will be indicated by $K$, while the mortgage rate that can be found in the market at time $t$ for a mortgage with maturity $T$ is denoted by,
\begin{equation}
   \kappa(t) \equiv \kappa(t;T,\zeta) := S_{t,T}(t) + \zeta,
\label{eqn:RefinancingBenchmark_Spread}
\end{equation}
where $\zeta$ denotes a deterministic spread (related to liquidity risk and the profit for the bank). This quantity is assumed to be constant over time and independent of the level of the interest rates. Note that the spread $\zeta$ will only be used for the pricing model. All calculations regarding hedging will be based on the model with $\zeta=0$, since a bank typically does not hedge the fixed coupon received by the mortgagor completely, but only the amount that corresponds to the funding costs.

Regarding the functional form of the incentive, there are two main approaches in the literature. The first one is based on the difference between the rates
\begin{equation}
    \epsilon(t) = K - \kappa(t),
    \label{eqn:RI_difference}
\end{equation}
which is the form that we will use (it has also been the model of choice in \cite{kolbe2008valuation, hoda2007implementation, perry2001study}). Clearly, the smaller the market rate, the greater the difference, with for an at-the-money mortgage  $\epsilon(t)=\epsilon^*=0$.
\begin{rem}
An alternative is using the ratio, i.e.,
$\epsilon(t) = \frac{K}{\kappa(t)}$, which is related to the evaluation of the present value of an annuity per Euro of the monthly payments, as defined by Equation (\ref{eqn:PresentValueAnnuity}). The ratio, $$\frac{\text{An}(t;\kappa(t))}{\text{An}(t_0;K)},$$ then provides a benchmark for the refinancing incentive. This ratio functional form has been introduced in \cite{richard1989prepayments} and it is used, among others, in~\cite{bandic2004pricing, van1998valuation}. Furthermore, in the literature, also the net present value gained by refinancing is sometimes used, see, for instance, \cite{jacobs2005modelling, svenstrup2002interest}.
\end{rem}
\begin{figure}[h]
\centering
\includegraphics[width=.47\textwidth]{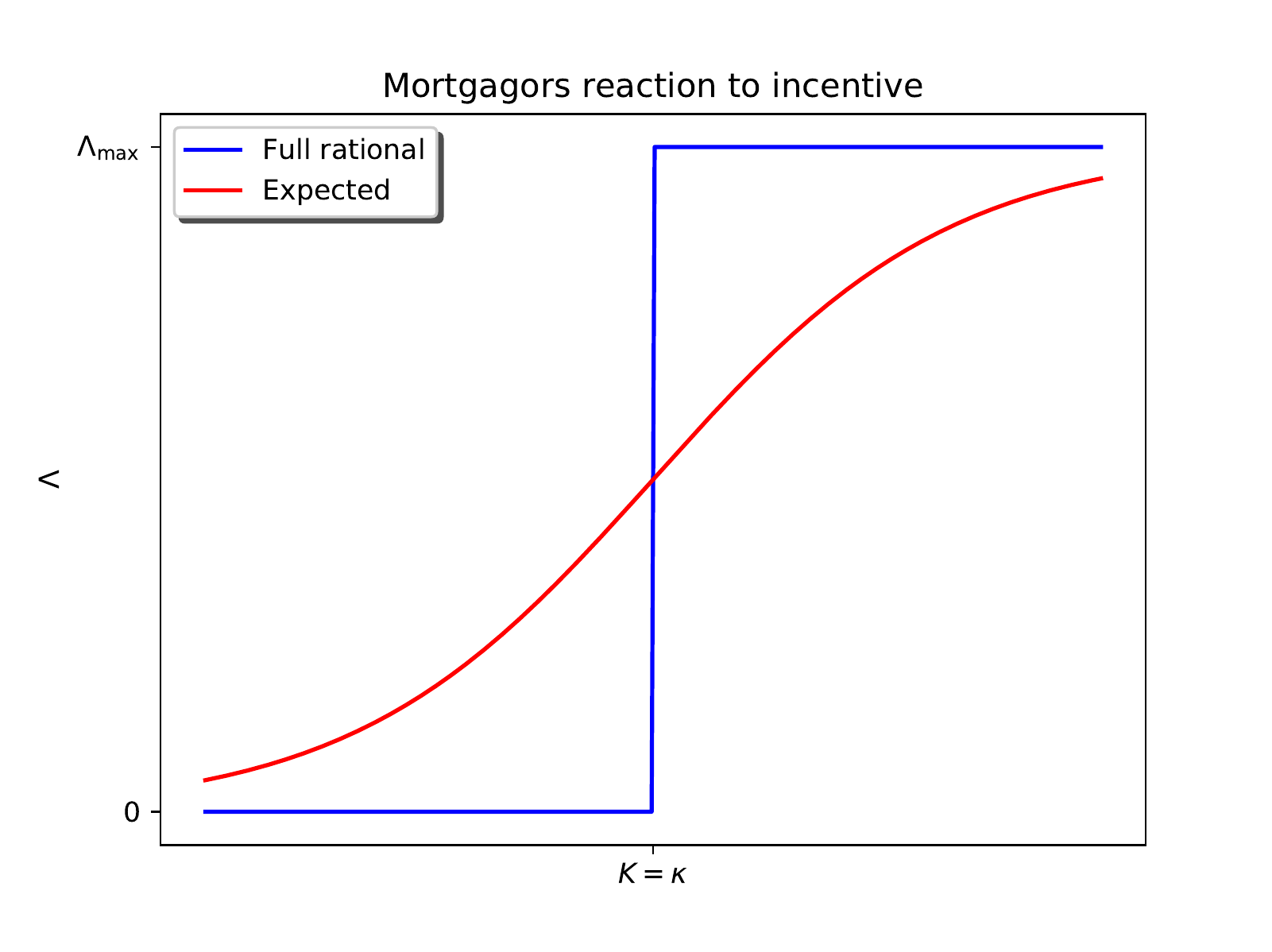}
\includegraphics[width=.47\textwidth]{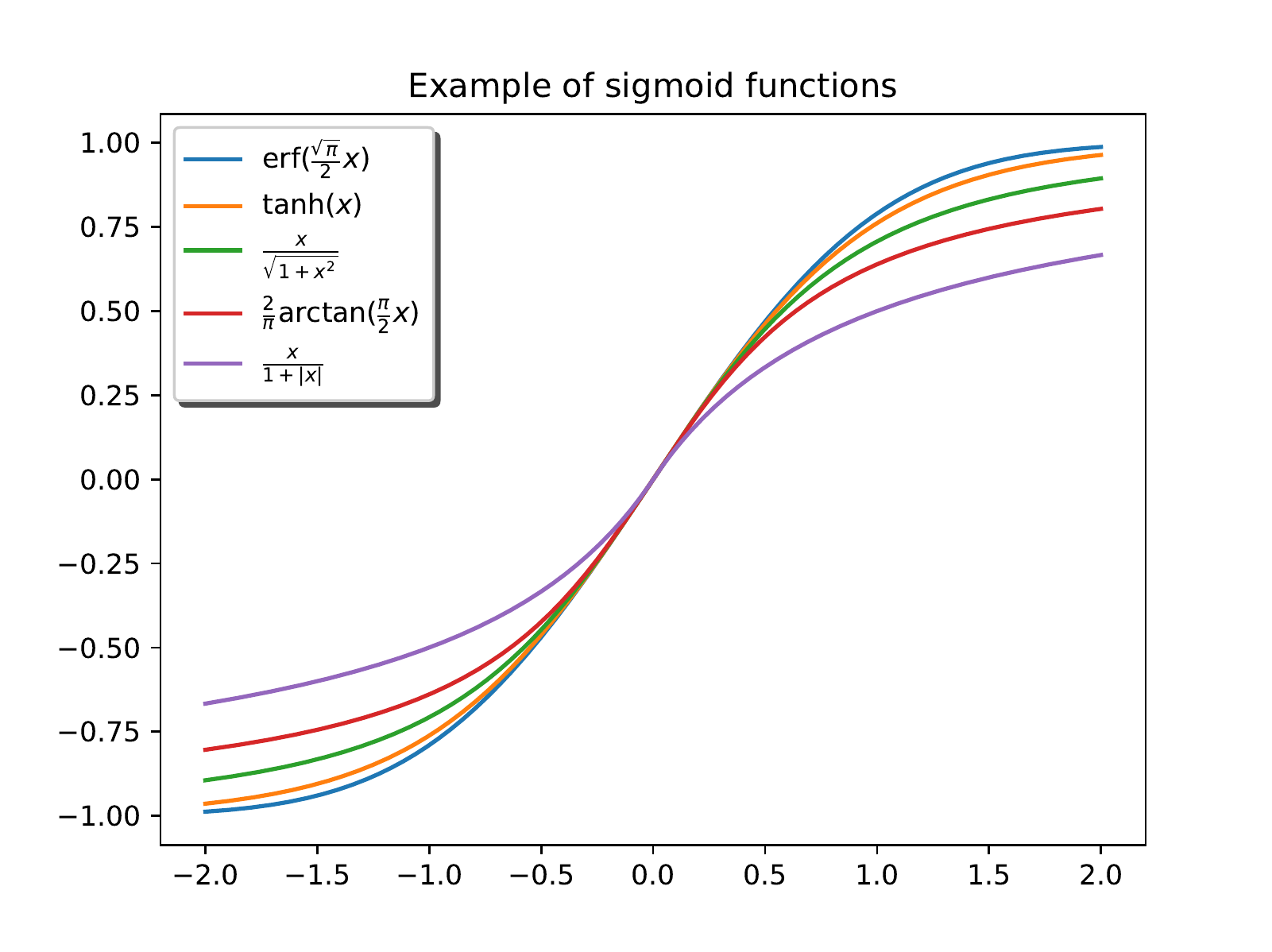}
\caption{Left: real versus expected reaction of the people to the refinancing incentive. Right: some examples of sigmoid functions, rescaled to have steepness one in the origin.}
\label{fig:incentive_behaviour}
\end{figure}
It is a well-known fact, that people do not always act rationally, meaning that they may not prepay when a natural opportunity arises, and mortgagors may prepay when it is not optimal. If rationality were a fact, the ``response function'' would be represented by the blue line in the left-side graph of Figure \ref{fig:incentive_behaviour}, where the maximum level of prepayments is reached, instantly, as soon as $\epsilon(t)>\epsilon^*$. Nevertheless, a smooth {\it $S$-shaped function}, such as the red line  in the figure, is a known model from the literature and appears to be more accurate than the step function to model prepayment behaviour. The $S$-shape  defines a class of functions, i.e., the ``sigmoid'' functions, where it is also possible to include a reaction time to the incentive, and some examples are presented in the right-side graph. For modeling the non-linear behaviour, an arc-tangent function has been widely applied \cite{bandic2004pricing,hoda2007implementation,castagna2013measuring,kolbe2008valuation}, but also a normal CDF \cite{hoda2007implementation, svenstrup2002interest}.

The analysis of more than thirty million rows of prepayment data resulted in a historical calibration of the refinancing incentive and  provided us with an approximation for the connection between the risk-neutral world of interest rates and the prepayment rate, for which a risk-neutral evaluation is not available. This calibration approach has a significant impact on the hedging strategy since evaluating the complete distribution of the notional unveils the necessity to include non-linear financial hedging instruments.

\section{Pricing Model}
\label{3}
In this section, we establish a link between the value of the mortgage portfolio and the level of the interest rates in the market (leaving aside the other possible drivers for prepayment).
Other prepayment drivers, that cannot easily be hedged as it is nontrivial to link them to instruments in the financial market, will be discussed in Section~\ref{da}. Their impact on the notional may be (partially) hedged by certain derivatives but this possible extension is left for further research. 

We develop a method to evaluate a portfolio of mortgages that takes the prepayment option embedded in such contracts into account. For this, we introduce a variation of the Index Amortizing Swap (IAS).
The main idea behind the pricing model is to replicate the mortgage portfolio by an IAS, whose notional will depend on the characteristics of the contract, the prepayment rate, and the interest rates in the market. Our model considers only the refinancing incentive as the driver of prepayments, but this will not lead to a fully rational model, because the smooth functional form for $\Lambda$ will be preferred over the step-function, see Figure~\ref{fig:incentive_behaviour}.




\subsection{Index Amortizing Swap \label{subsection:IAS}}
An IAS is an over-the-counter interest rate swap that combines a plain vanilla interest rate swap and, partially, a swaption. Amortizing swaps with a deterministic amortization scheme are commonly traded instruments. The peculiarity of the IAS which makes it ``a hybrid product'' is that its amortization scheme is predetermined only as a function of a specific interest rate. Therefore, the IAS is effectively acting as an option on that rate.

The dependency of the IAS on the interest rate level is the interesting aspect for prepayment replication.
An IAS and a mortgage portfolio essentially share the same embedded optionality. 
 From an interest rate risk perspective, only fixed-rate loans give rise to prepayment risk, because the loans with a variable rate will pay a coupon that is already adjusted to the market rate. The difference between the agreed mortgage rate and the at-the-market variable rate is then equal to zero because, effectively, $K(t) \equiv \kappa(t)$ at each time point $t$. So, fixed rate loans are usually hedged to reduce the  interest rate risk, and connecting mortgages with interest rate swaps achieves this.

\begin{figure}[h]
    \centering
    \includegraphics[scale=0.5]{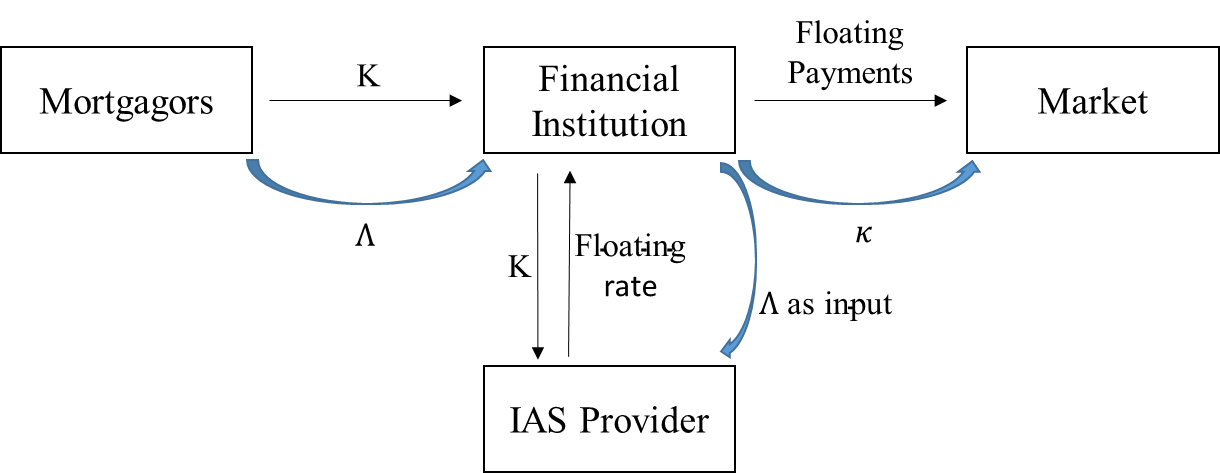}
    \caption{Representation how a bank would use an IAS to hedge prepayment risk, using prepayments as input to exchange the correct amount of fixed  rates for floating rates.}
    \label{fig:schemeIAS}
\end{figure}

However, interest rate swaps are not contractually linked to mortgage loans, and therefore prepayments may result in a misalignment of the cash flows of the hedge. In Figure~\ref{fig:schemeIAS}, the mechanism is schematically represented. The blue lines indicate prepayments, and the idea behind the IAS is to use the prepayment as an input, which then corrects the mismatch between the plain-vanilla swaps and the mortgage portfolio so that the precise fixed rate quantity can be exchanged for a floating one. When the floating rate has been received, it is reused to either fulfill some other floating rate commitments or to fund  new mortgages with an at-the-money market rate $\kappa$.

A few papers already established the link between Index Amortizing Swaps and portfolios of mortgages, for example, see \cite{galaif1993index, fernald1993pricing, jamshidian2005replication} or the textbook  \cite{rich1996modelling}, so the idea is not new. However, the construction presented in the following section defines an innovative approach to this problem. We link the observed market rates and the amortizing scheme of the IAS, taking into account the historical behaviour of mortgagors and the different repayment schemes of bullet or annuity mortgages. Moreover, we analyze the performance of hedging strategies that aim to replicate the non-linear payoff resulting from prepayments.

\subsection{Definition of the pricing model}
The starting point for the evaluation of the IAS is the following expectation,
\begin{equation}
    V_\text{IAS}(t_0) = \mathbbm{E}^\mathbbm{Q} \left[ \sum_{i=1}^{M} \tau_i \frac{N(T_{i-1};\Lambda(T_{i-1}))}{M(T_i)}   \cdot \big( K - L(T_{i-1};T_{i-1},T_i) \big) \Big | \mathcal{F}(t_0) \right],
    \label{eqn:IAS}
\end{equation}
where the IAS payment dates, $T_1,...,T_M$, are assumed to be the same as those of the mortgage portfolio. For simplicity, we assume yearly payments, so that $\tau_i \equiv 1,\;\; \forall i$. Moreover, the notional of the IAS at time $T_i$ is assumed to be the notional of the mortgage at time $T_{i-1}$. Also note that the prepayment rate $\Lambda(T_i)$ is here one of the dependent arguments in the notional $N$, and $K$ is the mortgage rate which is the at-the-money rate at time $t_0$. Discounting is denoted by $M(\cdot)$, and $L(T_{i-1};T_{i-1},T_i)$ is the Libor rate, with trade date $T_{i-1}$, start date $T_{i-1}$ and maturity date $T_{i}$, defined as,
\[L(t;T_{i-1},T_i)=\frac{1}{\tau_i}\frac{P(t,T_{i-1})-P(t,T_{i})}{P(t,T_{i})},\;\;\;\tau_i=T_{i-1}-T_i,\]
and where $P(t,T_i)$ are the zero coupon bonds with maturity $T_i.$

An aspect to take into account is the type of mortgage. In Sections~\ref{sec:bullet} and~\ref{sec:annuity}, closed-form expressions for the values of the annuity and bullet mortgage plans were presented, based on a constant prepayment rate $\Lambda$. Generally, we need to explicitly state the dependency on the repayment scheme at each time. Here, we use,
\begin{equation}
    N(T_i) = N(T_{i-1}) \cdot \Psi(\Lambda(T_i)),
    \label{eqn:NotionalDependencyPsiFunction}
\end{equation}
where function $\Psi(\cdot)$ is defined as follows:
\begin{equation}
\Psi =
\begin{cases}
1 - \Lambda(T_i) & \text{(Bullet),} \\[1.0ex]
1 + \frac{\displaystyle K(\Lambda(T_i) - 1)}{\displaystyle 1 - (1+K)^{-(T_M-T_{i-1})}} + K - \Lambda(T_i) \cdot (K+1)  & \text{(Annuity)},
\end{cases}
\label{eqn:MortgageTypeFunctionPsi}
\end{equation}
with $i=1,...,M$. Using $N(T_0)=N_0$, Equation (\ref{eqn:NotionalDependencyPsiFunction}) is well-defined.

A second dependence relates to the modeling of the prepayment rate $\Lambda(t)$. In practice, it is common to adopt a constant  $\Lambda$-value for the whole year, which is empirically inaccurate, as confirmed by the corresponding hedging strategy. Such a model neglects the non-linear part of the prepayment risk.  An enhanced feature here is the inclusion of randomness in $\Lambda$. The literature is scarce regarding stochastic models for the notional of a mortgage portfolio. One of the few approaches based on a stochastic process to forecast the prepayment rate is found in \cite{ahmad2018stochastic}, which provides a rigorous model based on individual mortgagors. Estimating the drift and volatility of each borrower's wealth process for a vast portfolio of mortgages has, however, feasibility drawbacks, and the model's applicability is therefore reduced. On the other hand, strictly empirical models do not offer the insight to distinguish the interest rate incentive from  other variables, making the implementation of a stochastic extension for the modeling of the notional challenging. The assumption we make here is that the prepayment rate is only determined by the refinancing incentive (denoted by RI), i.e.,
$
    \Lambda(T_i) = \Lambda(\RI(T_i)).
$

Subsequently, we need to choose a functional form for the refinancing incentive, referring to Figure \ref{fig:incentive_behaviour}, with a ``fully rational'' exercise function and the alternative model. Choosing a smooth function for RI aims to capture the empirical behaviour of mortgagors and is connected to the fact that prepayments are not guided by only one variable (which, in theory, would lead to the fully rational model). Two functional forms will be compared in the forthcoming hedging experiments to validate the numerical procedure and analyze real-world scenarios. So, RI is given in the following two forms:
\begin{equation}
\RI(T_i) =
\begin{cases}
    \Lambda_{\max} \mathbbm{1}_{ \left \{ \epsilon(T_i) > \epsilon^* \right\} } & \text{(fully rational model)}, \\[1.0ex]
    \alpha_1 + \alpha_2 \left(1 + \e^{\alpha_3 \epsilon(T_i) + \alpha_4} \right)^{-1} & \text{(alternative S-shaped model)}.
\end{cases}
\label{eqn:RefinancingIncentiveFunctionalForm}
\end{equation}
The logistic function which is often used in the literature, has been chosen to model human behaviour here, however, there is no apparent reason to prefer it over other S-shaped functions.

A final modeling step is regarding $\epsilon(\cdot)$, which models the relation between the mortgage rate $K$ and the market rate $\kappa(t)$. The authors of~\cite{alink2002mortgage} have evaluated different models based on prepayment data of Dutch mortgages and concluded that the best results wre obtained by the model
$\epsilon(T_i) = K - \kappa(T_i)$, see also~(\ref{eqn:RI_difference}), which will also be used here.

All dependencies in the IAS notional are now motivated.
The following equations summarise the model formulation for the evaluation of a (Dutch) mortgage portfolio, in compact form:


\begin{equation}
\boxed{
\begin{aligned}
V_{\text{IAS}}(t_0) &=& \mathbbm{E}^\mathbbm{Q} \left[ \displaystyle \sum_{i=1}^{M} \tau_i \frac{\displaystyle N(T_{i-1})}{\displaystyle M(T_i)} \left( K - L(T_{i-1};T_{i-1},T_i) \right) \bigg | \mathcal{F}(t_0) \right] \hspace*{0.5cm} \text{\small Index Amortizing Swap,}  \\
N(T_i) &=& N(T_{i-1}) \cdot \Psi(\Lambda(T_i)) \hspace*{6.0cm} \text{\small Mortgage type } (\ref{eqn:MortgageTypeFunctionPsi}),   \\[1.0ex]
\Lambda(T_i) &=& \RI(T_i;\epsilon(T_i)) \hspace*{6.5cm} \text{\small Prepayment driver } (\ref{eqn:RefinancingIncentiveFunctionalForm}),  \\[1.0ex]
\epsilon(T_i) &=& K - \kappa(T_i) \hspace*{7.7cm}  \text{\small Market conditions}. \end{aligned}
}
\label{eqn:final_system}
\end{equation}

\begin{rem}[Historical calibration of the refinancing incentive]
\label{subsection:DataAnalysis}
One of the main drawbacks of any prepayment model is that prepayments need to be calibrated to historical data because there are no traded instruments that give the information about risk-neutral prices of prepayment options. Our approach to approximating a risk-neutral valuation of a mortgage portfolio consists of two stages. First of all, we calibrate the interest rate model and the volatilities observed in the market. Secondly, the prepayments are linked to the interest rates' market level by a historical calibration of the mortgagor's reactions to the interest rate incentives. This way, the functional form of the refinancing incentive connects the risk-free levels of the interest rates to the forecasted/predicted prepayment rates.
\end{rem}

\subsection{Data analysis}\label{da}
An important aim is to forecast/assess the {\it Conditional Prepayment Rate} (CPR). An early definition of the CPR is found in \cite{fabozzi1992mortgage}.

\begin{definition}[Conditional Prepayment Rate]
The CPR is an annualised rate of prepayment, obtained from a measure which is called the Single Monthly Mortality (SMM), and it is based on the following formulas:
\begin{eqnarray}
\SMM(t) & =& \frac{ \displaystyle{\textrm{Unscheduled notional payment at month }} t}{\displaystyle \text{Scheduled outstanding at month } t}, \\[1.0ex]
\CPR(t) & =& 1 - (1 - \SMM(t))^{12}.
    \label{eqn:DefinitionCPR}
\end{eqnarray}
\end{definition}
When considering an entire portfolio of mortgages, which is our primary interest, the CPR is frequently observable. When we refer to a ``mortgage'', we thus mean a ``portfolio of mortgages''.
\begin{figure}[h]
    \centering
    \includegraphics[scale=0.3]{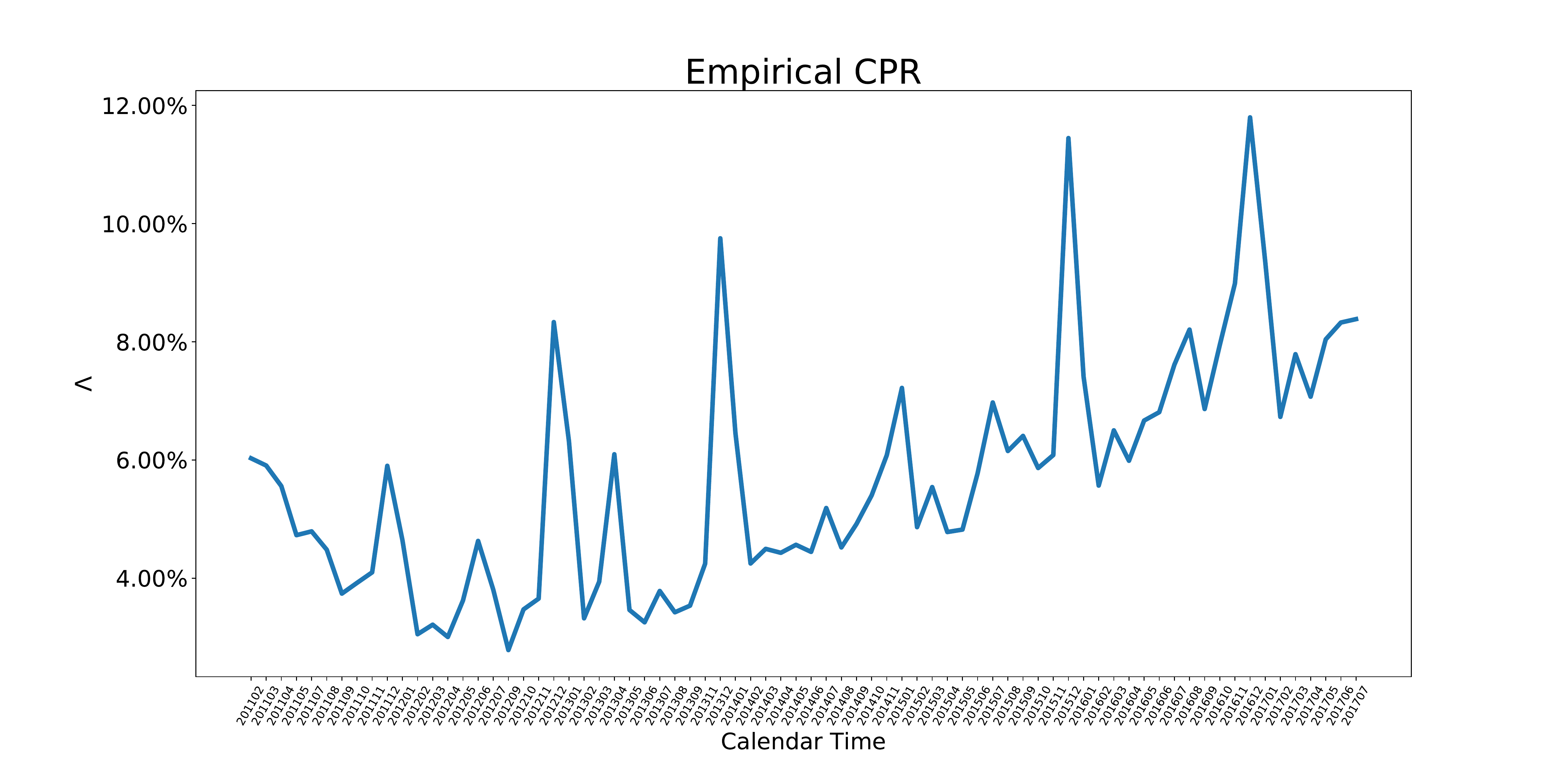}
    \caption{Empirical CPR obtained from more than 31 million observations collected in a time span of 74 months.}
    \label{fig:EmpiricalCPR}
\end{figure}

The reasons behind prepayment are diverse. Usually, in the Netherlands, ten or twenty percent of the notional can be prepaid each year without paying a penalty. This is a significant difference between the Dutch and the US markets, where, in the latter market, usually, the notional can be paid back entirely at any time.  Partial prepayments are also referred to as \textit{curtailments}. Penalties do not always apply in the Netherlands, as most banks offer the possibility of full debt prepayment without extra costs if the mortgagor relocates. In periods of economic growth and/or when the housing market is booming, prepayments due to movement occur therefore frequently.

Here, the focus is on the fixed-rate personal mortgages that represent the largest segment, with a notional of more than 150 billion Euro. The data set in use accounts for monthly  observations of, on average, more than 400 000 mortgages over 74 months, from February 2011 to July 2017, resulting in a total of more than 30 000 000 observations. The variables used include,
\begin{itemize}
\setlength\itemsep{0.0em}
    \item ``StartingBalance'', the notional of the mortgage at the end of a month, which only considers contractual repayments.
    \item ``PrepaidAmount'', the magnitude of a prepayment.
    \item ``Period'', the month and year of the observation.
    \item ``InterestRateIncentive''. There are different interest rate incentives in the data set, depending on the rate used. Imagine a mortgage of ten years, observed after three years. One may argue that the market rate represents either the incentive for a ten-year mortgage or a seven-year mortgage. In agreement with the pricing model developed, the second option is chosen here.
\end{itemize}
With the significant variables defined, using the definition of prepayment rate (\ref{eqn:DefinitionCPR}), we extrapolate the empirical CPR from the data set, evaluating
\begin{equation}
    \SMM(\text{Period}_i) = \frac{ \sum_{j=1}^{N_i} \text{PrepaidAmount}_{i,j} }{ \sum_{j=1}^{N_i} \text{StartingBalance}_{i,j} }, \;\;\; \CPR(\text{Period}_i) = 1 - (1 - \SMM(\text{Period}_i))^{12},
\label{eqn:DefinitionEmpiricalCPR}
\end{equation}
where $N_i$ represents the number of mortgages in the ``Period'' (month) $i$. Figure~\ref{fig:EmpiricalCPR} presents the empirical prepayment rate for each of the 74 months, based on Equation (\ref{eqn:DefinitionEmpiricalCPR}). $\Lambda(t)$ increases on average, which may be attributed to the very low (and even negative) levels of the interest rates that characterised the market since 2011, leading to a high prepayment incentive. The observed peaks in the figure are related to the fact that mortgagors tend to prepay more, in certain months, as extra money due to salary bonuses becomes available.
\begin{figure}[h]
    \centering
    \includegraphics[scale=0.65]{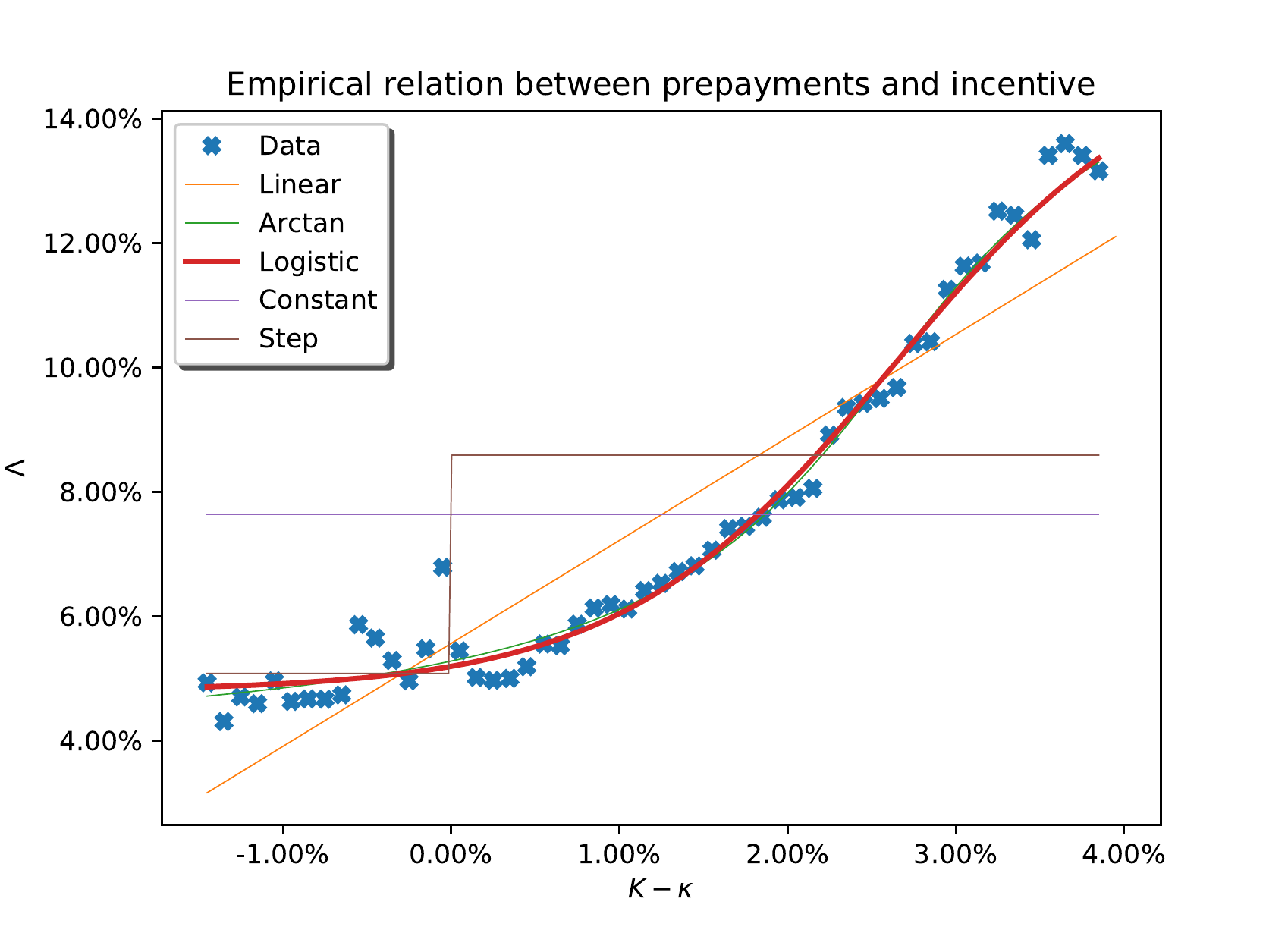}
    \caption{Prepayment data show an empirical, non-linear relation between the level of prepayment and the incentive $K - \kappa$, where $K$ indicates the "old" mortgage rate and $\kappa$ is the at-the-money market rate when prepayment occurs.}
    \label{fig:RealPrepaymentData_Interpolations}
\end{figure}
In the refinancing incentive, our analysis does not take the timing of the prepayments into account. The data is collected  in ``bins of mortgages'' that share the same refinancing incentive. This is essential as mortgages observed in a certain period $i$ may not share the same maturity, rate, incentive, and there may even be different contracts (annuities or bullet) characteristics. Collecting these mortgages of a specific month in an ``averaged mortgage'', with an averaged rate, maturity, and averaged incentive would result in a loss of important information. We circumvent this by a mortgage-focused adaptation, for which we define,
\begin{equation*}
    \SMM_m := \frac{ \text{PrepaidAmount}_{m} }{ \text{StartingBalance}_{m} },
\end{equation*}
which represents the monthly prepayment rate for each observation $m$, with $m=1,...,30 000 000$. We concentrate on the range of incentives, $K-\kappa = [-1.5\% ; 4\%]$, as almost all observations fall into this interval. The range is subdivided into 56 equally-sized bins. Prepayment rate $\SMM_m$ ends up in bin $b$, if the incentive in  row $m$ belongs to the interval covered by bin $b$. Resulting are quantities $\SMM_{m,b}$, where the subscripts indicate the index of the observation and the bin to which it belongs. Moreover,
\begin{equation*}
\begin{split}
    \overline{\SMM}_b  = \frac{1}{N_b} \sum_{m=1}^{N_b} \SMM_{m,b},\;\;\;     \CPR_b  = 1 - (1 - \overline{\SMM}_b))^{12},
\end{split}
\end{equation*}
with $N_b$ the number of observations in bin $b$. The results are shown in Figure \ref{fig:RealPrepaymentData_Interpolations}, and they exhibit a pattern. Different functions have been ``fitted'', using least-squares regression. A constant prepayment rate is found to be inaccurate and gives a misinterpretation of the relation between prepayments and market movements. Models assuming a fully rational behaviour of mortgagors also lack precision since a step function model does not return accurate results. The sigmoid functions plotted in Figure \ref{fig:RealPrepaymentData_Interpolations}, appear accurate representations (where the logistic function is chosen).


\subsection{IAS evaluation with deterministic CPR function}\label{4}
Equation (\ref{eqn:IAS}) forms the basis for the evaluation of the IAS. We start with a basic test case, assuming that the prepayment rate is a {\it deterministic} function of time, $\Lambda(t).$ This simplification implies that we only need to analyze two ``levels'' of the model, i.e.,
\begin{equation*}
\begin{cases}
V_{\text{AS}}(t_0) = \mathbbm{E}^\mathbbm{Q} \left[ \displaystyle \sum_{i=1}^{M} \tau_i \frac{\displaystyle N(T_{i-1})}{\displaystyle M(T_i)} \left( K - L(T_{i-1};T_{i-1},T_i) \right) \bigg | \mathcal{F}(t_0)  \right] & \text{\small Index Amortizing Swap,} \\[2.5ex]
N(T_i) = N(T_{i-1}) \cdot \Psi(\Lambda(T_i)) & \text{\small Mortgage type } (\ref{eqn:MortgageTypeFunctionPsi}).
\end{cases}
\end{equation*}
Because there is no need to define the prepayment rate further, the IAS is just an amortizing swap (AS). In this setting, we will have an analytic solution for the price of the AS, which is helpful. Because the solution will be obtained without a specific interest rate model, it provides us with an easy check. When a short-rate model is employed, and $\Lambda$ is (still) considered deterministic and independent of the market, the price of the IAS from the simulation of bonds and Libor rates should return the same price for the AS. Furthermore, this deterministic case can be interpreted as a model in which  forecasting the prepayment rate is performed separately, for example, based on an in-depth data analysis or artificial intelligence.

The price of the AS, with the time-dependent notional, is given by,
\begin{equation}
\begin{split}
    V_\text{AS}(t_0) & = \mathbbm{E}^\mathbbm{Q} \left[\sum_{i=1}^{M} \tau_i \frac{N(T_{i-1};\Lambda(T_{i-1}))}{M(T_i)} \cdot \big( K - L(T_{i-1};T_{i-1},T_i) \big) \big | \mathcal{F}(t_0) \right] \\
    & = \sum_{i=1}^{M} \tau_i P(t_0,T_i) N(T_{i-1}) \left( K - \mathbbm{E}^{T_i} \left[ L(T_{i-1};T_{i-1},T_i) \big | \mathcal{F}(t_0) \right] \right) \\
    & = \sum_{i=1}^{M}  N(T_{i-1}) \left[ P(t_0,T_i) (\tau_i K + 1) - P(t_0,T_{i-1}) \right].
\end{split}
\label{eqn:PriceDeterministicIAS}
\end{equation}

\subsection{IAS with stochastic CPR \label{subsection:IASwithStochasticCPR}}
We focus on stochastic interest rates and their inclusion in the amortization schedule, as presented in the Equations~(\ref{eqn:final_system}).
In~\cite{richard1989prepayments} it was explained that, in the case of a deterministic refinancing rate, the forecasted prepayments cannot be accurate. The deterministic setting won't give us any insight in the distribution of the notional of the IAS, as only one path is considered. For this reason, the generalization towards a stochastic environment is essential and valuable.

In our experiments, we start with $\kappa(T_i)$ to arrive at $N(T_i)$. The quantities that need to be simulated are the discount factor $M(T_i)$, the Libor rate $L(T_{i-1}; T_{i-1}, T_{i})$ and the market mortgage rate $\kappa(T_i)$. In the simulation, we prefer a short-rate model for $r(t)$. Among the short-rate models, the Hull-White and CIR++ models will be employed because the prices of zero-coupon bonds and European options on bonds and swaptions can be found analytically ~\cite{OosterleeGrzelakBook}.

A stochastic generalization of $\Lambda$ is straightforward, since all steps are based on the evolution of the short-rate $r(t)$.
A very common process, especially including negative interest rates, is the Hull-White SDE model, i.e.,
\[\d r(t) =\lambda\left(\theta(t)-r(t)\right)\dt+\eta \dW_r(t),\]
where $\theta(t)$ is a time-dependent drift term, which is used to fit the mathematical bond prices to the yield curve observed in the market and $W_r(t)\equiv W_r^\Q(t)$ is the Brownian motion under measure $\Q$. Parameter $\eta$ determines the overall
level of the volatility and $\lambda$ is the reversion rate parameter.
A large $\lambda$ value causes short-term rate movements to dampen rapidly, so that long-term volatility is reduced.
Parameter $\theta(t)$ is defined as, \begin{eqnarray*}\theta(t)=\frac{1}{\lambda}\frac{\partial }{\partial
t}f(0,t)+f(0,t)+\frac{\sigma^2}{2\lambda^2}\left(1-\e^{-2\lambda
t}\right).\end{eqnarray*}

With Equations~(\ref{eqn:RefinancingBenchmark_Spread}) and~(\ref{eqn:RI_difference}), the prepayment incentive, $\epsilon(t)$, is defined as \[\epsilon(t)=K-S_{t,T}(t)-\zeta,\]
where $S_{t,T}(t)$ is the swap rate and where $\zeta$ denotes a deterministic spread. The swap rate $S_{t,T}(t)$ is defined as,~\cite{OosterleeGrzelakBook},
\begin{eqnarray}
\label{eqn:SwapHW}
S_{T_m,T_n}(t)=\frac{P(t,T_{m})-P(t,T_n)}{\sum_{k=m+1}^n\tau_k
P(t,T_k)},\;\;\text{with}\;\;\;\tau_k=T_k-T_{k-1}.
\end{eqnarray}
Because of the affine structure of the Hull-White model, the zero-coupon bonds $P(t,T)$ are known explicitly,
$P(t,T)=\exp\left({A(\tau)+B(\tau)r(t)}\right)$,
where functions $A(\tau)$ and $B(\tau)$ are given by:
\begin{eqnarray*}
A(\tau)= -\frac{\eta^2}{4\lambda^3}\left(3+\e^{-2\lambda\tau}-4\e^{-\lambda\tau}-2\lambda\tau\right)+\lambda\int_0^\tau\theta(T-z) B(z)\d z,\;\;\;B(\tau)= -\frac{1}{\lambda}\left(1-\e^{-\lambda \tau}\right).
\end{eqnarray*}

Interest rate paths for $r(t)$ are then simulated, so that the zero-coupon bonds $P(T_i,T_j)$ can be approximated. With these, the swap rate $S_{T_m,T_n}(t)$ in~(\ref{eqn:SwapHW}) is determined, and subsequently the IAS in System~(\ref{eqn:final_system}) can be evaluated.



\begin{figure}[h]
    \centering
    \includegraphics[width=0.495\textwidth]{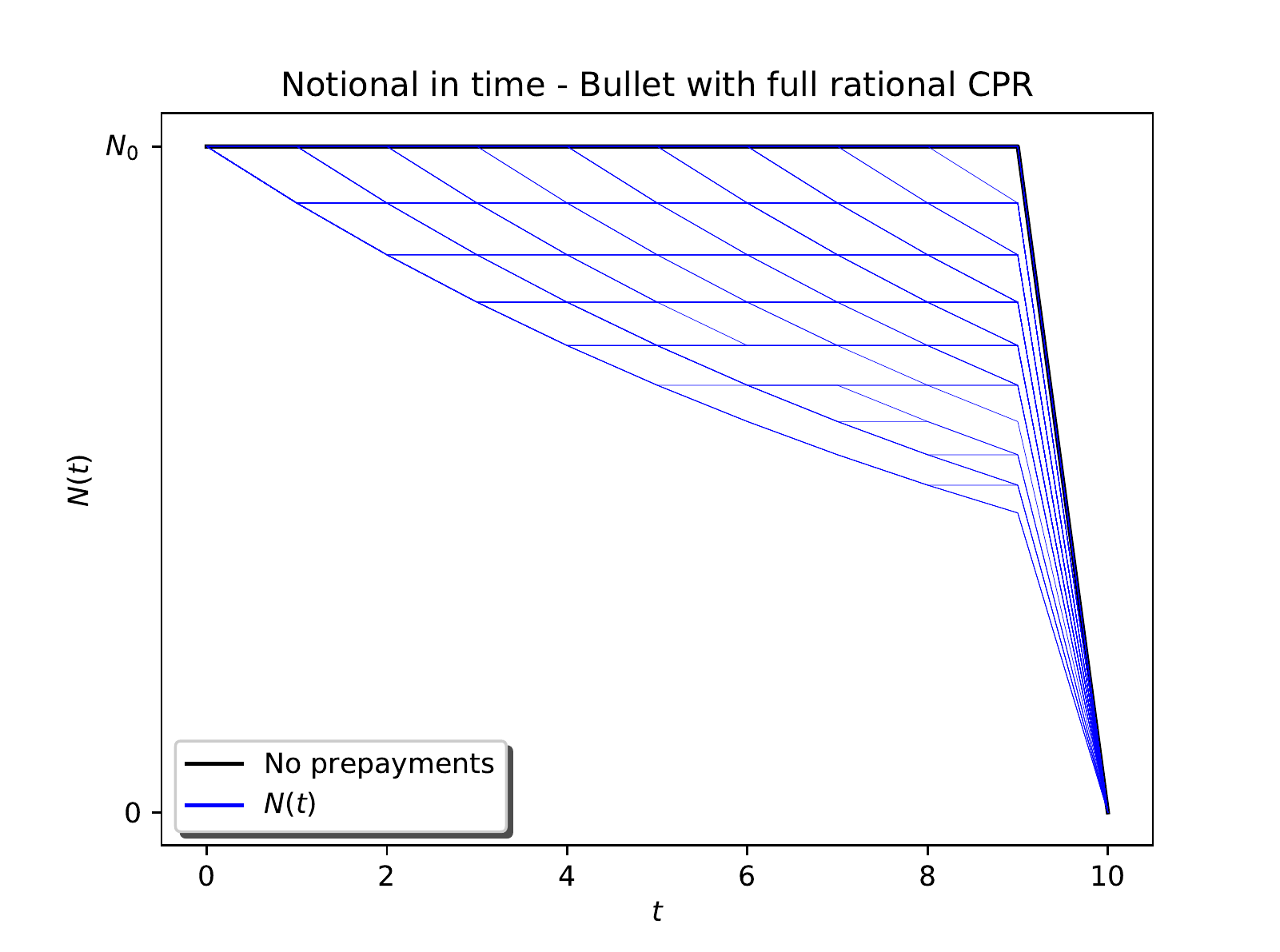} \hfill
    \includegraphics[width=0.495\textwidth]{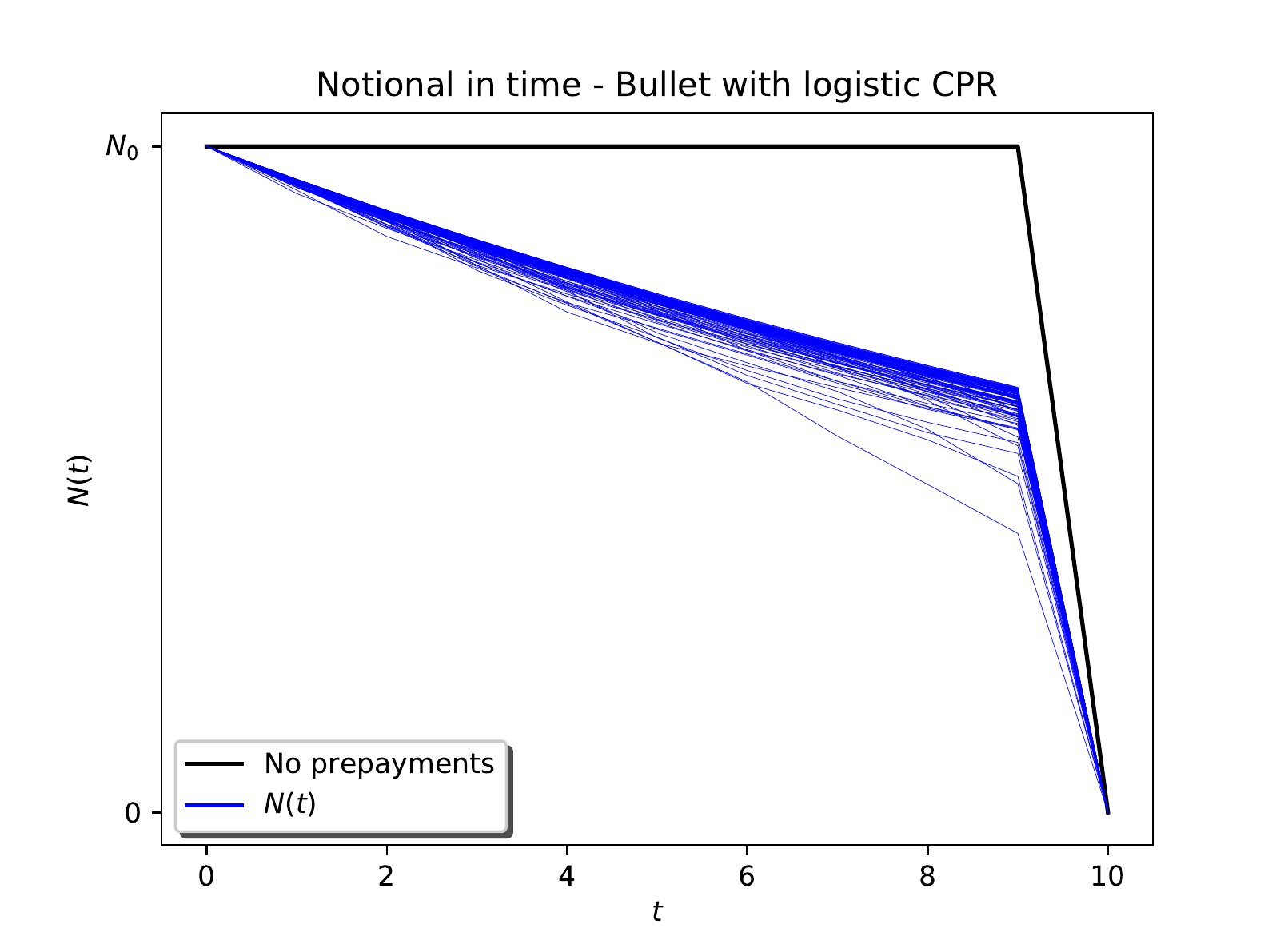} \\
    \includegraphics[width=0.495\textwidth]{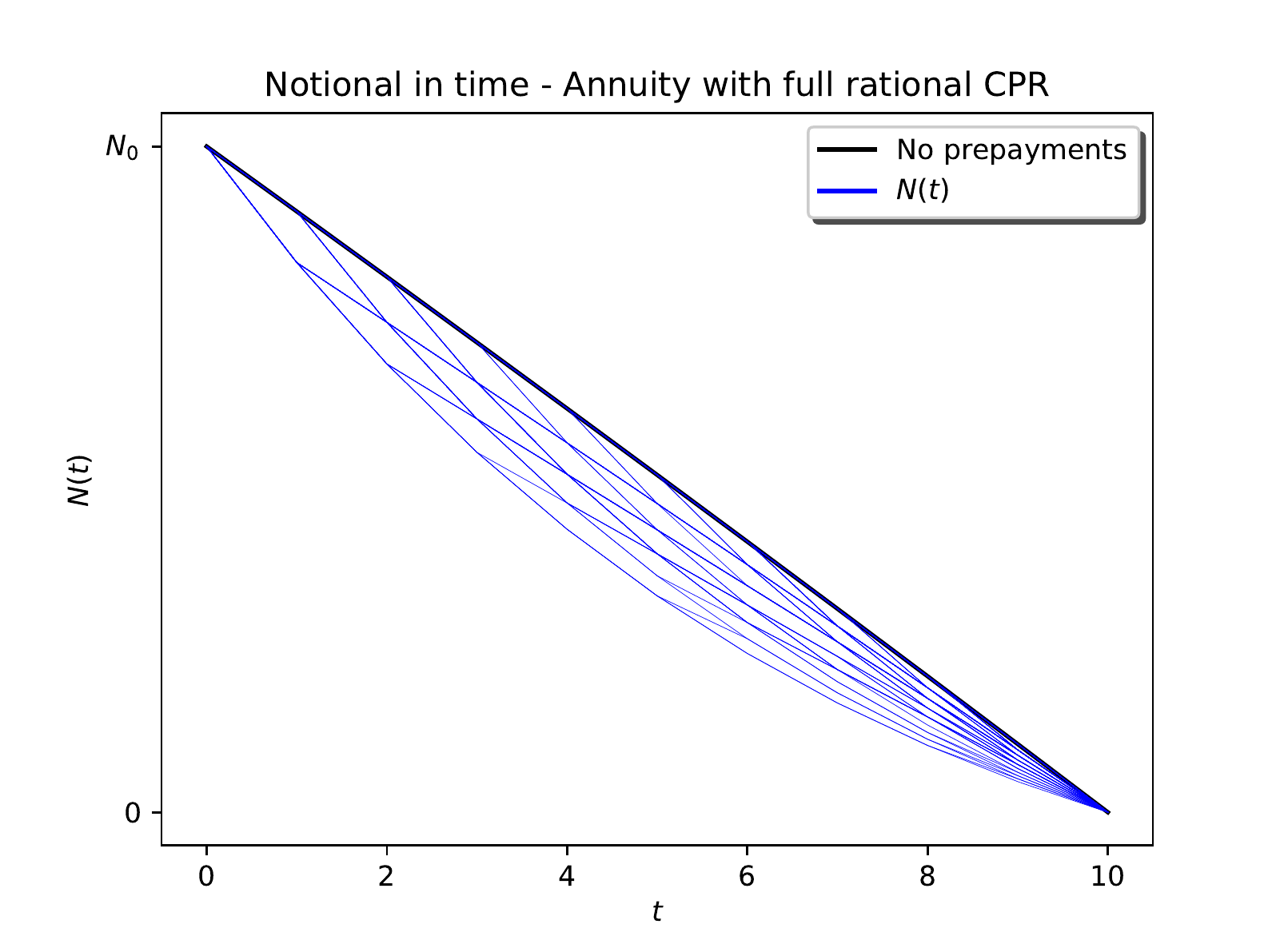} \hfill
    \includegraphics[width=0.495\textwidth]{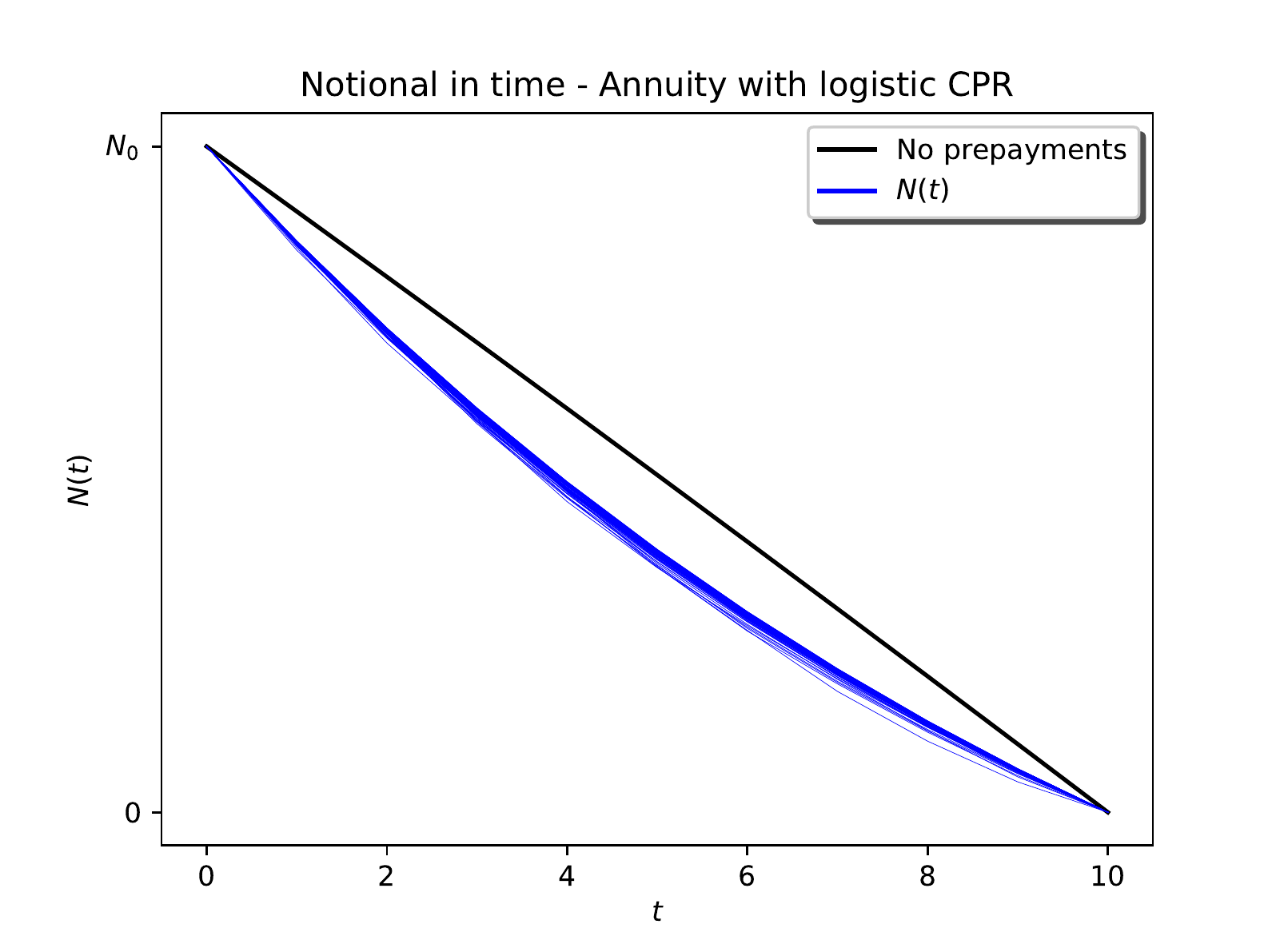}
\caption{Examples of the evolution of stochastic notionals for different contracts. First row: Bullet mortgage plan. Second row: Annuity mortgage plan. First column: Fully rational prepayment model. Second column: The prepayments follow the logistic function calibrated to historical data.}
    \label{fig:StochasticNotionalFirstExamples}
\end{figure}

Figure \ref{fig:StochasticNotionalFirstExamples} illustrates different examples of the notional $N(t)$, based on different types of contracts, comparing the refinancing incentive in the form of a step function with the sigmoid function. For completeness, the figure is related to Table \ref{tbl:NumericalResults:StochasticNotional}, which presents the details of the four cases considered.

\begin{table}
\centering
\footnotesize
\begin{tabular}{ccccc}
\toprule
Mortgage Type & $T_M$ & $K$ & $\Lambda$ & Numerical Price \\
 & (Years) & (bps) &  & (bps)\\
\midrule
Bullet & $10$ & $88.83$ & Full-rational & $-63.02$ \\
Bullet & $10$ & $88.83$ & Sigmoid & $74.93$ \\
Annuity & $10$ & $52.93$ & Full-rational & $-24.09$ \\
Annuity & $10$ & $52.93$ & Sigmoid & $36.28$ \\
\bottomrule
\end{tabular}
\caption{Values corresponding to the graphs of Figure \ref{fig:StochasticNotionalFirstExamples}.}
\label{tbl:NumericalResults:StochasticNotional}
\end{table}

\subsection{Discussion of the results}
Figure~\ref{fig:StochasticNotionalFirstExamples} reveals interesting features that need to be discussed.
The results of the simulations with the fully rational prepayment model, which was based on the maximum level of prepayment $\Lambda_{\text{max}}$ as soon as there is a prepayment incentive, exhibit a grid-like structure for the notional over time. This is consistent with the insight that, at each time, the mortgagors choose to either continue with the scheduled amortization plan or to prepay a fixed portion of the outstanding notional. However, since $\Lambda$ in (\ref{eqn:DefinitionCPR}) is proportional to the outstanding notional, we do not see fixed steps in the resulting grid (because significant amortization is reached for higher $N(t)$-values). Thus, at the top and left regions of the grid, we observe bigger jumps, while at the bottom and right side, the grid appears to be finer. This implies that, independent of the magnitude of $\Lambda_{\text{max}}$, early termination of the loan is not possible in this formulation~\footnote{Early termination could be included by a redefinition of the prepayment rate as a quantity proportional to the initial notional $N_0$.}. The insight that an early contract termination is not possible makes sense when the focus is indeed on a portfolio of mortgages.
Even if the grid does not have a stochastic component, the actual distribution of the $N(t)$-values at each point in time is stochastic.

Focussing on the experiments with the S-shaped refinancing incentive, i.e., the figures in the second column of Figure \ref{fig:StochasticNotionalFirstExamples}, in the top figure, an artificial black path is added to indicate how $N(t)$ would look if prepayment would not occur. The cloud of blue paths separates the  graphs in essentially three different regions. Below the black line and above $N(t)$, there is an area where prepayments are supposed to always happen, regardless of the rationality of the action. This is attributed to the calibrated logistic function, which is  positive and thus implies a minimum amount of prepayments, regardless of the market conditions. On the other hand, below the paths of the notional, there is an area where prepayments are supposed to never take place, despite the incentive. This is because the S-shaped function also presents a maximum.

Table \ref{tbl:NumericalResults:StochasticNotional} shows the results of the numerical simulations, presenting the IAS price in basis points. When the refinancing incentive is modeled as the step function, a negative price is displayed, which is consistent with the fully rational exercise strategy. If a mortgagor exercises the prepayment option, which is only driven by financial rationality, all opportunities to save money would be employed and the prepayments would lead to a loss for the bank. However, if the refinancing incentive is based on non-rational exercise and a delayed reaction to an incentive, a positive value will result.


\section{Hedging Strategies}
\label{5}
It is difficult (or expensive) to achieve  a perfect resemblance of the derivative characteristics with a hedging portfolio in practice. Often, a subset of the instruments is selected, and the position is dynamically adjusted to keep the values of the hedge position and the derivative sufficiently close.

The hedge that we will build is ``static'', which means that the risk is ideally addressed by a portfolio of tradeable instruments that need not be recalibrated in the future. The static replication may provide insight into the risks embedded in the dynamics of a mortgage portfolio.

A common way to hedge prepayment risk is by utilizing swaps. However, this may give an inaccurate approximation of the IAS price and the Greeks,  under all possible scenarios.
We will focus on the non-linear risk generated by the prepayment option and also the Greeks are calculated.
 The difference between a linear and non-linear hedging strategy will be discussed. Although the choice for a static hedging strategy may be considered a limitation, it forms the basis for a dynamic hedging strategy.


\subsection{Linear hedge strategy}
\begin{figure}
    \centering
    \includegraphics[scale=0.55]{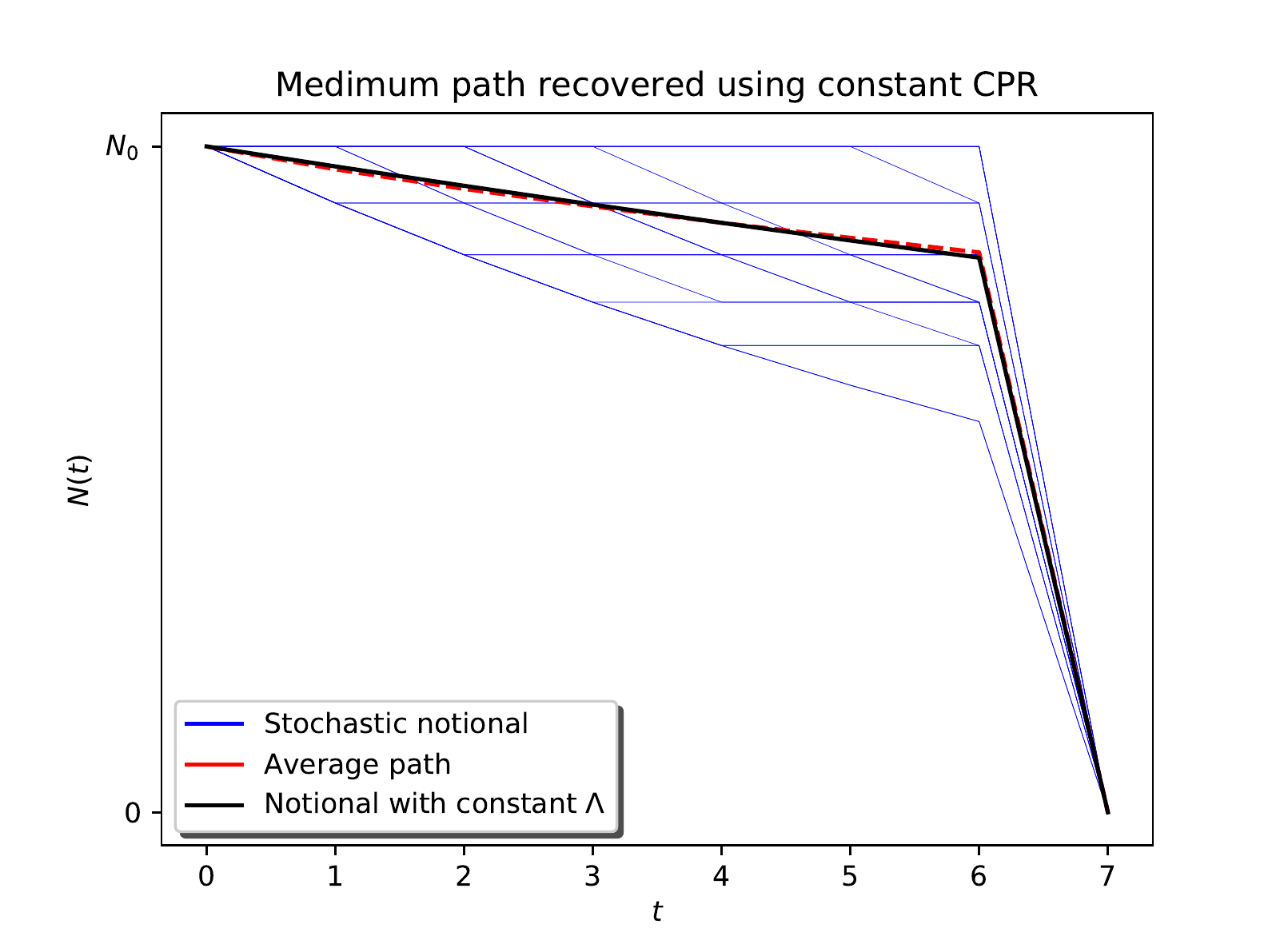}
    \caption{Example of the average path recovered by using a constant prepayment rate. This result can be obtained regardless of the functional form of $\Lambda$. Here, the full-rational model has been used (so that the paths of $N(t)$ do not overlap).}
    \label{fig:MediumPathRecoveredWithConstantCPR}
\end{figure}

A linear hedge implies that a movement in the underlying asset price will affect the present value of the instrument in linear way. The most common example of a linear hedge instrument is the plain-vanilla swap (or simply ``the swap''). Using a linear hedge, we can replicate one path and thus focus only on the average path. The following definition of a (receiver) swap, with notional value 1, a delayed start at time $T_l$ and an anticipated end time $T_m$, will be used,
\begin{equation*}
    V_{\text{S}}(t_0;l,m) = \sum_{i = l + 1}^m P(t_0,T_i) \tau_i \big(K - L(t_0;T_{i-1},T_i) \big).
\end{equation*}
Then, an AS is a linear combination of co-terminal\footnote{A decomposition in co-initial standard swaps is also possible.} standard swaps,
\begin{equation*}
    V_{\text{AS}}(t_0) = \sum_{i = 1}^M  \overline{\Delta N}(T_{i-1}) V_S(t_0;i,M),
\end{equation*}
where  $\overline{\Delta N}(T_i) := \overline{N}(T_i) - \overline{N}(T_{i-1})$, and $\overline{N}(T_{-1}) = 0$. To define an accurate amortization, we use the average notional of the IAS, at each point in time, as
\begin{equation*}
    \overline{N}_{\text{IAS}}(T_i) =\E[\overline{N}_{\text{IAS}}(T_i)]= \frac{1}{N_{\text{Sim}}}\sum_{j = 1}^{N_{\text{Sim}}} N^{(j)}_{\text{IAS}}(T_i).
\end{equation*}
The value of the linear hedge is then given by
\begin{equation}
    V_{\text{AS}}(t_0) = \sum_{i = 1}^M \overline{\Delta N}_{\text{IAS}}(T_{i-1}) V_S(t_0;i,M).
\label{eqn:ValueLinearHedge}
\end{equation}
Notice that the average path of $N_{\text{IAS}}(t)$ can be attributed to a constant prepayment rate. In fact, we can determine the constant $\Lambda$-value which recovers the average path of $N_{\text{IAS}}$, as presented in Figure \ref{fig:MediumPathRecoveredWithConstantCPR}. This does not mean that the price of the IAS is the same as the price of the swap with a deterministic amortization schedule which resembles the average path of the stochastic notional, because, in general, we cannot assume that the distribution of the notional is independent of the distribution of the Libor rate, meaning that,
\begin{equation*}
\begin{split}
    \mathbbm{E}^{T_i} \left[ N_{\text{IAS}}(T_{i-1}) \big( K - L(T_{i-1}) \big)  \right] \neq \mathbbm{E}^{T_i} \left[ N_{\text{IAS}}(T_{i-1}) \right] \cdot  \mathbbm{E}^{T_i} \left[ \big( K - L(T_{i-1}) \big)  \right],
\end{split}
\end{equation*}
with $L(T_{i-1}):=L(T_{i-1};T_{i-1},T_i).$


\subsection{Non-linear hedge strategy}
\begin{figure}[h]
    \centering
    \includegraphics[width=0.495\textwidth]{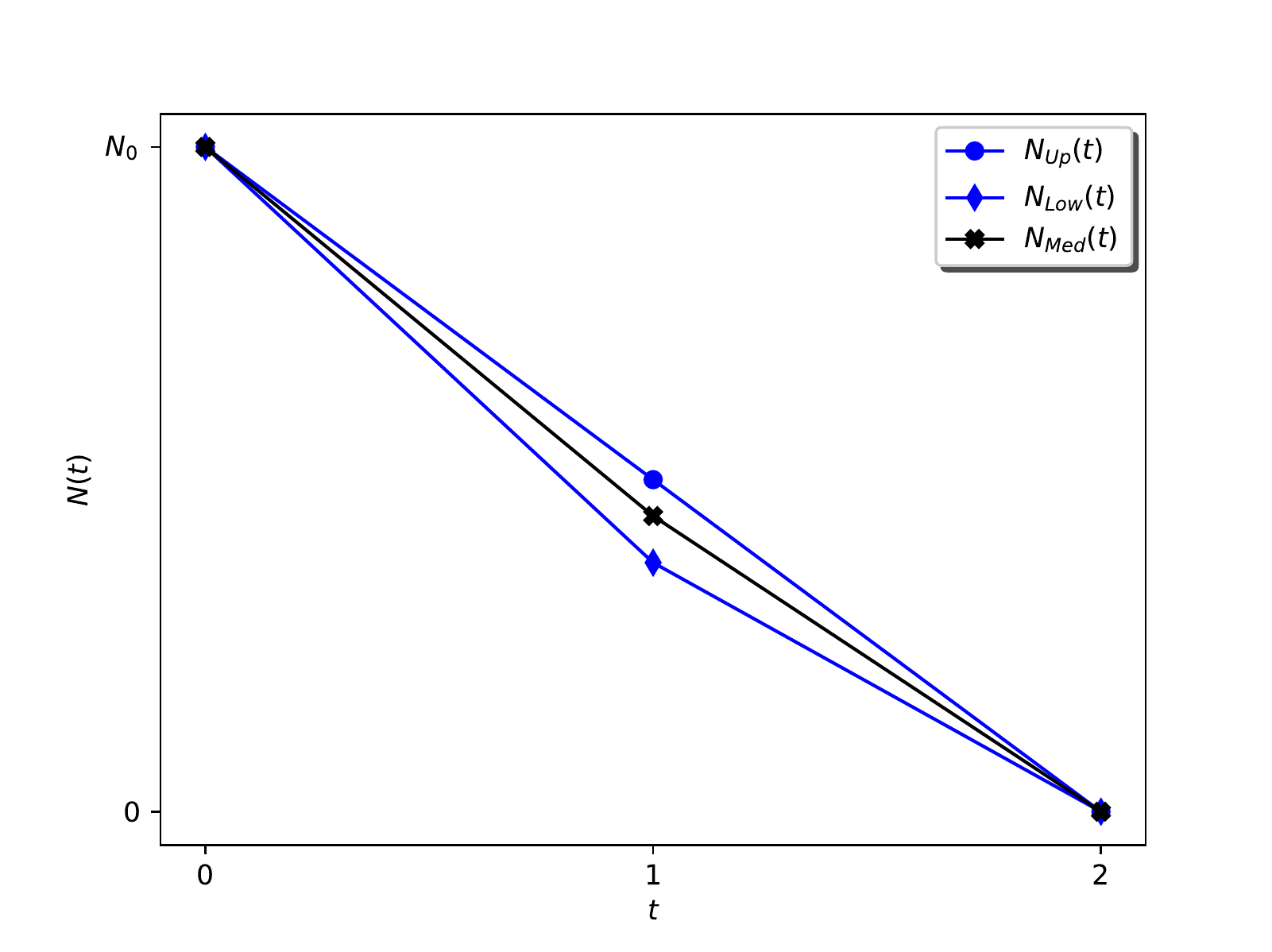} \hfill
    \includegraphics[width=0.495\textwidth]{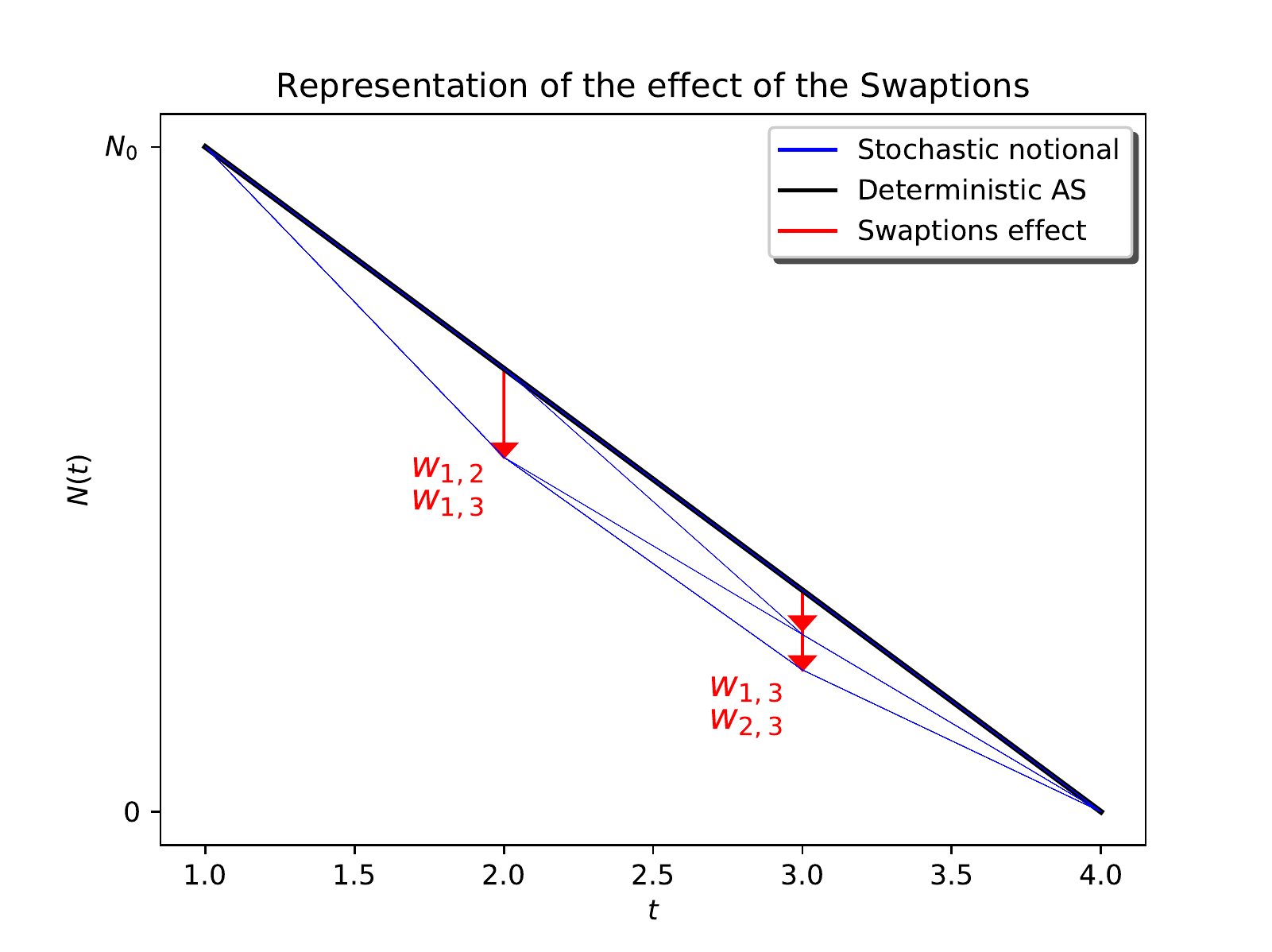}
    \caption{Left: the toy model which leads to an analytical price of the IAS. Right: A combination of swaptions with notionals $w_{i,j}$ reduces the notional of the hedging portfolio, replicating the (blue colored) notional of the amortizing swap. 
    }
    \label{fig:ToyModelAndSwaptionJustification}
\end{figure}

We will use non-linear financial instruments to achieve an improved hedge performance.  Consider an annuity with initial notional $N(0) = N_0$, maturity time $T_M = 2$ and the fixed rate $K$. Moreover, assume we will use the  fully rational functional form for the refinancing incentive, as in (\ref{eqn:RefinancingIncentiveFunctionalForm}). This model example gives us a closed-form solution for the price of the IAS. We start from the expectation formulation, with $M=2$, i.e.,
\begin{equation*}
    V_{\text{IAS}}(t_0) = \mathbbm{E}^\mathbbm{Q} \left[ \displaystyle \sum_{i=1}^{2} \tau_i \frac{N_{\text{IAS}}(T_{i-1})}{M(T_i)} \left( K - L(T_{i-1};T_{i-1},T_i) \right) \big | \mathcal{F}(t_0) \right].
\end{equation*}
The only possible prepayment may take place at $T_2$, so the randomness is in $N(T_1)$, and the first payment is deterministic, i.e.,
\begin{equation*}
\begin{split}
    C(T_1) = P(t_0,T_1) N_{\text{IAS}}(t_0) \big(K - L(t_{0};t_{0},T_1) \big).
\end{split}
\end{equation*}
Since in this example, the prepayment rate equals the step-function value, we can assess the possible notional values at time $T_1$, being $N^{\text{Up}}$ or $N^{\text{Low}}$, as shown in Figure \ref{fig:ToyModelAndSwaptionJustification}. Furthermore, using the notation $L(T_1):=L(T_1;T_1,T_2)$, it follows that $S_{T_1,T_2}(T_1) = L(T_1)$,
so that the notional at time $T_1$ can be written as
\begin{equation*}
\begin{split}
N(T_1) & = N^{\text{Up}} \mathbbm{1}_{ \{ K < L(T_1) \} } + N^{\text{Low}} \mathbbm{1}_{ \{ K > L(T_1) \} } \\
& = N^{\text{Up}} - \left( N^{\text{Up}} - N^{\text{Low}} \right) \mathbbm{1}_{ \{ K > L(T_1) \} }.
\end{split}
\end{equation*}
The second payment $C(T_2)$ is therefore equal to,
\begin{equation*}
\begin{split}
    C(T_2) & = M(t_0) \mathbbm{E}^\mathbbm{Q} \left[ \frac{N(T_{1})}{M(T_2)} \tau_2 \big(K - L(T_{1}) \big)  \big | \mathcal{F}(t_0) \right]  \\
    & = \mathbbm{E}^\mathbbm{Q} \left[ \frac{M(t_0)}{M(T_2)} N^{\text{Up}} \big(K - L(T_{1}) \big)  \big | \mathcal{F}(t_0) \right] - 				\mathbbm{E}^\mathbbm{Q} \left[ \frac{M(t_0)}{M(T_2)} \left(N^{\text{Up}} - N^{\text{Low}}\right) \big(K - L(T_{1}) \big)^+  \big | \mathcal{F}(t_0) \right] \\[1.0ex]
    & = N^{\text{Up}} \big((K + 1)P(t_0,T_2) - P(t_0,T_1) \big) - \left(N^{\text{Up}} - N^{\text{Low}}\right) V_{\text{Floorlet}}(t_0;T_1,T_2).
\end{split}
\end{equation*}
With all payments in an explicit form, the price of the instrument is found to be,
\begin{eqnarray}
 V_{\text{IAS}}(t_0) & = & C(T_1) + C(T_2)  \nonumber \\
& = &\sum_{i=1}^2 N(T_i) \big((K + 1)P(t_0,T_i) - P(t_0,T_{i-1}) \big) - \left(N^{\text{Up}} - N^{\text{Low}}\right) V_{\text{Floorlet}}(t_0;T_1,T_2)\nonumber \\[1.0ex]
& =& V_{\text{AS}} (t_0) - \left(N^{\text{Up}} - N^{\text{Low}}\right) V_{\text{Floorlet}}(t_0;T_1,T_2).
\label{eqn:PriceToyModelAnnuity}
\end{eqnarray}
This price defines the IAS as a combination of an Amortizing Swap and a Floorlet. This is an interesting insight, because we can now separate the linear component and a non-linear one, suggesting that  a long position in swaps plus a short position in an option on the refinancing incentive replicates the IAS in a more accurate way.
Purchasing a mortgage can thus be seen as entering a long position in a swap, while the prepayments effectively reduce the notional of the mortgage. Thus, the option to prepay would be equal to an option on a swap, i.e., a swaption.

\subsection{Portfolio construction \label{subsection:PortfolioConstruction}}
The combination of swaps replicates an Amortizing Swap, and the main insight is using swaptions to reduce the notional of the AS in the same way prepayments will reduce the notional of the IAS. An assumption is that the swaps and swaptions have the same payment frequency and the same fixed and floating rates as the IAS. A mismatch between the cash flows of the IAS and the portfolio of swaps and swaptions will then only be caused by the difference in the notionals.

This static, non-linear hedge strategy aims to offset the behaviour of the IAS \textit{in a path-wise fashion}. IAS simulated paths are required for the calibration of the portfolio. So, an ``a-posteriori calibration'' should take place.
An essential reference for this is \cite{jamshidian2005replication}, where the authors defined a model-free calibration of a portfolio of Bermudan swaptions to replicate a so-called flexi-swap. The amortization of their IAS however involved bounds on the amortization, and they assumed that the notional either continued without amortization or jumped from the upper bound to the lower bound. We cannot use their model in our framework, as modelling any ``half-redemption'' would be impossible, whereas we emphasize that, historically, mortgagors do not act in a rational way. Other essential references include \cite{joshi2001pricing,carr1998static}. Additionally, in~\cite{JableckiMortgages}, a methodology for estimating the value of a prepayment option in an illiquid market in interest rate swaptions is discussed.

We will use swaps to replicate the upper bound on the paths of the IAS, representing the minimum amount of prepayment. Considering that a mortgage has payment dates $T_1,...,T_M$ potentially all the swaptions presented in Table \ref{tbl:AllPossibleSwaptions} are available, where $w_{i,j}$ indicates the notional of a swaption, starting at time $T_i$ and ending at time $T_j$. Figure \ref{fig:ToyModelAndSwaptionJustification}, right figure, summarizes the concepts expressed so far.
\begin{table}
\centering
\begin{tabular}{lllll}
\toprule
& & Swaptions & & \\
\midrule
$w_{1,2}$ & $w_{1,3}$ & $w_{1,4}$ & $\cdots$ & $w_{1,M}$ \\
$-$       & $w_{2,3}$ & $w_{2,4}$ & $\cdots$ & $w_{2,M}$ \\
$\vdots$  & $-$       & $\ddots$  &          & $\vdots$ \\
          & $\vdots$  &           &          &   \\
$-$       & $-$       & $\cdots$  &          & $w_{M-1,M}$ \\
\bottomrule
\end{tabular}
\caption{The possible swaptions available to hedge the prepayment risk. Element $w_{i,j}$ in the table indicates the notional of the swaption with value $V_{\text{Swp}}(t_0;T_i,T_j)$.}
\label{tbl:AllPossibleSwaptions}
\end{table}

Mathematically, we describe the portfolio as:
\begin{equation}
    \Pi(t_0,\mathbf{w}) = \underbrace{V_{\text{AS}}(t_0,K)}_{\text{Long swaps}} \underbrace{- \sum_{i=1}^{M-1} \sum_{l=i+1}^{M} w_{i,l} V_{\text{Swp}}(t_0;T_i,T_l)}_{\text{Short receiver swaptions}},
\label{eqn:PortfolioDef_LongAS_ShortSwaptions}
\end{equation}
and the problem is to determine the set of weights, representing the notionals of the swaptions, such that,
\begin{equation}
    \textbf{w}^* = \argmin_{\textbf{w}} F(\textbf{w}),
\label{eqn:MinimizationFunctionWeightsSwaptionGeneral}
\end{equation}
where $F(\textbf{w})$ is a function that measures the  ``distance'' between $N_{\text{IAS}}$ and $N_{\Pi}$. More specifically, the value of $N_\Pi(T_k;\textbf{w})$ in the $j$-th simulation  is given by,
\begin{equation}
\begin{split}
    N_\Pi^{(j)}(T_k;\textbf{w}) & = N_{\text{AS}}(T_k) - \sum_{i=1}^{k-1} \sum_{l=i+1}^{M} w_{i,l} \mathbbm{1}_{\{ K>\kappa(T_i) \}}^{(j)}, \\
    N_{\text{AS}}(T_k) & = \max_j N_{\text{IAS}}^{(j)}(T_k),
\end{split}
\label{eqn:DefinitionNotionalHedgingPortfolio}
\end{equation}
where the indicator function models whether or not the prepayment is triggered. Notice that $N_{\text{AS}}(t)$ does not depend on $j$, because it is deterministic. Depending on the form of $F(\textbf{w})$ in~(\ref{eqn:MinimizationFunctionWeightsSwaptionGeneral}), different quantities will be minimized, and the optimal solution may be different too. A reasonable choice for $F(\textbf{w})$ is:
\begin{equation}
    F(\textbf{w}) = \sum_{\text{Time } t} \frac{1}{N_{Sim}} \sum_{j}^{N_{Sim}} \left[ \left | N_{\text{IAS}}^{(j)}(t) - N_\Pi^{(j)}(t;\textbf{w}) \right |^2 \right].
\label{eqn:MinimizationFunction_SumTimeAverageSimulations}
\end{equation}
as a mismatch between the notional of the IAS and of $\Pi(t,\mathbf{w})$ is included in each scenario $j$, and the squared errors are averaged over the simulations and integrated over time.

We will set up a ``minimum-variance'' hedge. A useful result is an analytic minimization formula for a particular set of swaptions, which is illustrated in the following proposition.
\begin{figure}[h]
    \centering
    \includegraphics[scale=0.55]{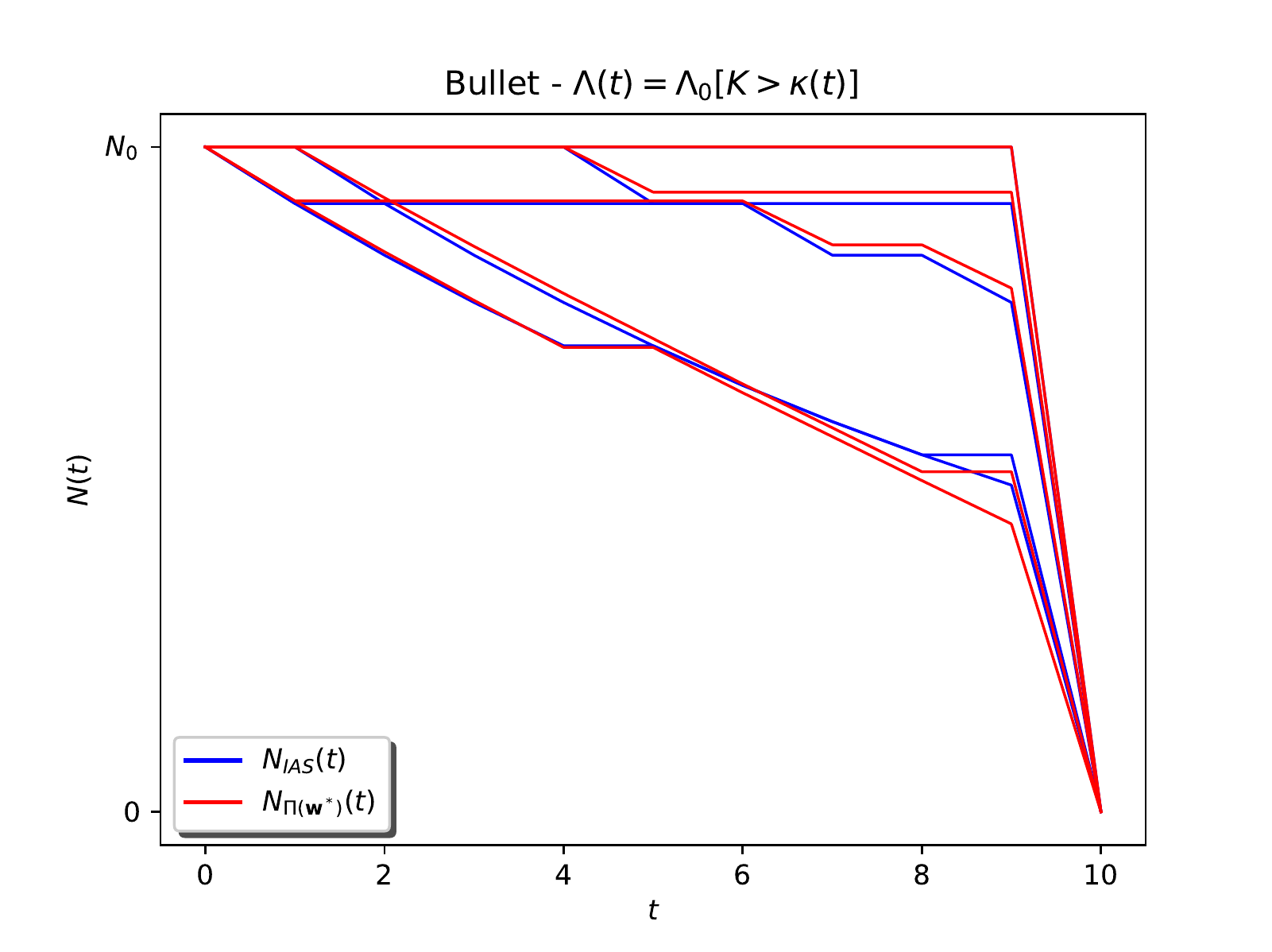}
    \caption{Example how a calibrated portfolio may replicate the paths of the notional of the IAS. The case with the fully rational prepayment rule has been chosen because the paths are spread out, so it is possible to visualize the replication.}
    \label{fig:HedgingExample_BulletFewPaths}
\end{figure}
\begin{proposition}[Calibration of the diagonal swaptions on the paths of the IAS]
Consider the swaptions on the diagonal in Table \ref{tbl:AllPossibleSwaptions}, i.e., the swaptions whose start date and tenor sum up to the maturity of the mortgage. For example, for a mortgage with maturity $T_M=10$, take the swaptions with notional $w_{1,10}, w_{2,10}, ... , w_{9,10}$. Define for simplicity $    \mathbbm{1}_{i}^{(j)} = \mathbbm{1}_{\{ K > \kappa^{(j)}(T_i) \}}.$
The solution to problem (\ref{eqn:MinimizationFunctionWeightsSwaptionGeneral}), using the function (\ref{eqn:MinimizationFunction_SumTimeAverageSimulations}), corresponds to solving
\begin{equation*}
    \nabla F(\textbf{w}) = \left[ \frac{\partial F}{\partial w_{1,M}} \ , \ ... \ , \ \frac{\partial F}{\partial w_{M-1,M}} \right] = \textbf{0}.
\end{equation*}
This requires the solution of the linear system $A \mathbf{w} = \mathbf{b}$. Particularly,
\begin{equation}
\begin{split}
A & =
\begin{bmatrix}
     (M-1) \sum_{j=1}^{N_{Sim}}  \mathbbm{1}_{1}^{(j)} \mathbbm{1}_{1}^{(j)} & \dots &
     (M-M+1) \sum_{j=1}^{N_{Sim}}  \mathbbm{1}_{1}^{(j)} \mathbbm{1}_{M-1}^{(j)} \\
     (M-2) \sum_{j=1}^{N_{Sim}}  \mathbbm{1}_{2}^{(j)} \mathbbm{1}_{1}^{(j)} & \dots &
     (M-M+1) \sum_{j=1}^{N_{Sim}}  \mathbbm{1}_{2}^{(j)} \mathbbm{1}_{M-1}^{(j)} \\
     \vdots & \vdots & \vdots  \\
     (M-M+1) \sum_{j=1}^{N_{Sim}}  \mathbbm{1}_{M-1}^{(j)} \mathbbm{1}_{1}^{(j)} & \dots &
     (M-M+1) \sum_{j=1}^{N_{Sim}}  \mathbbm{1}_{M-1}^{(j)} \mathbbm{1}_{M-1}^{(j)}
\end{bmatrix},
\\[1.5ex]
\mathbf{b} & =
\begin{bmatrix}
     \sum_{k=2}^M \sum_{j=1}^{N_{Sim}} \mathbbm{1}_{1}^{(j)} \left( \overline{N}(T_k) -  N_{\text{IAS}}^{(j)}(T_k) \right) \\
     \sum_{k=3}^M \sum_{j=1}^{N_{Sim}}  \mathbbm{1}_{2}^{(j)} \left( \overline{N}(T_k) -  N_{\text{IAS}}^{(j)}(T_k) \right) \\
     \vdots \\
     \sum_{k=M-1}^M  \mathbbm{1}_{M-1}^{(j)} \left( \sum_{j=1}^{N_{Sim}} \overline{N}(T_k) -  N_{\text{IAS}}^{(j)}(T_k) \right),
\end{bmatrix}
\\[1.5ex]
\mathbf{w} & =
\begin{bmatrix}
     w_{1,M} & w_{2,M} & \dots & w_{M-1,M}
\end{bmatrix}^T.
\end{split}
\label{eqn:SolutionCalibrationHedgingPortfolio}
\end{equation}
\label{proposition:DiagonalSwaptionsCalibration}
\end{proposition}

This proposition states that the calibration using the swaptions on the diagonal in Table \ref{tbl:AllPossibleSwaptions} can be performed without any numerical method. For other sets of swaptions, similar results can be found. Note that these swaptions correspond to the counter-diagonal, for example, in Table \ref{tbl:SwaptionVolatilityMatrix}.

\begin{figure}[h]
    \centering
    \includegraphics[width=0.495\textwidth]{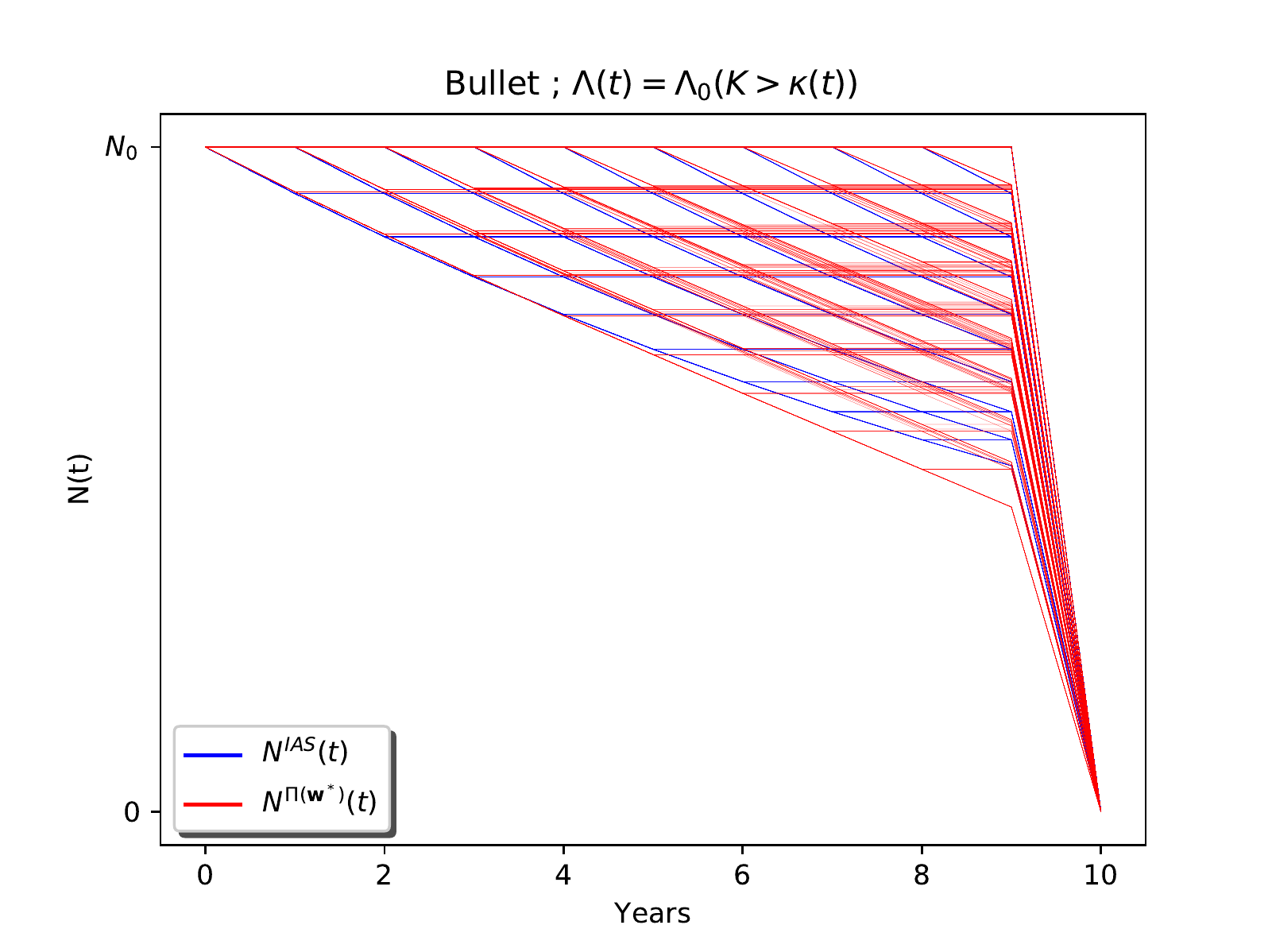} \hfill
    \includegraphics[width=0.495\textwidth]{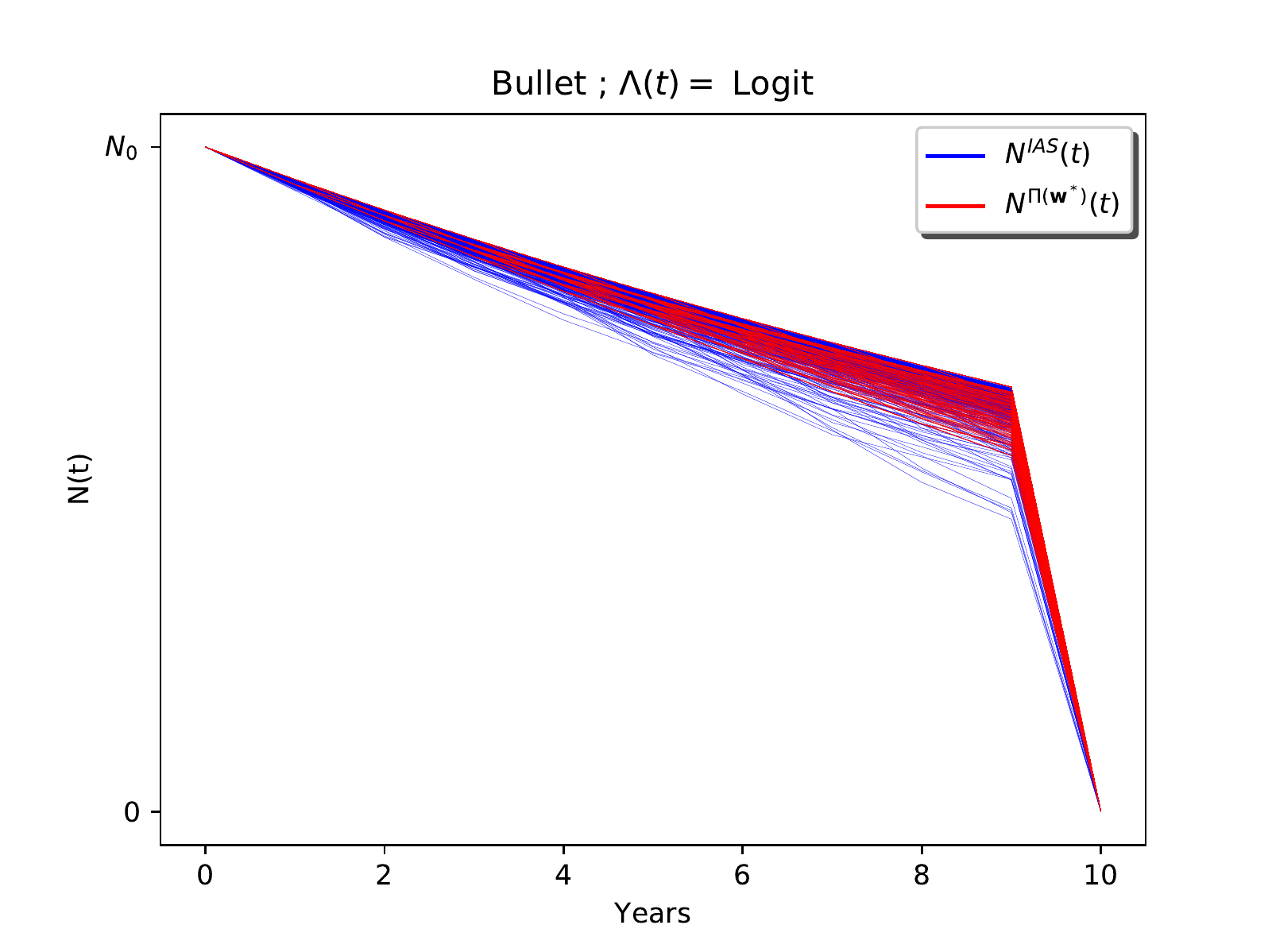} \\
    \includegraphics[width=0.495\textwidth]{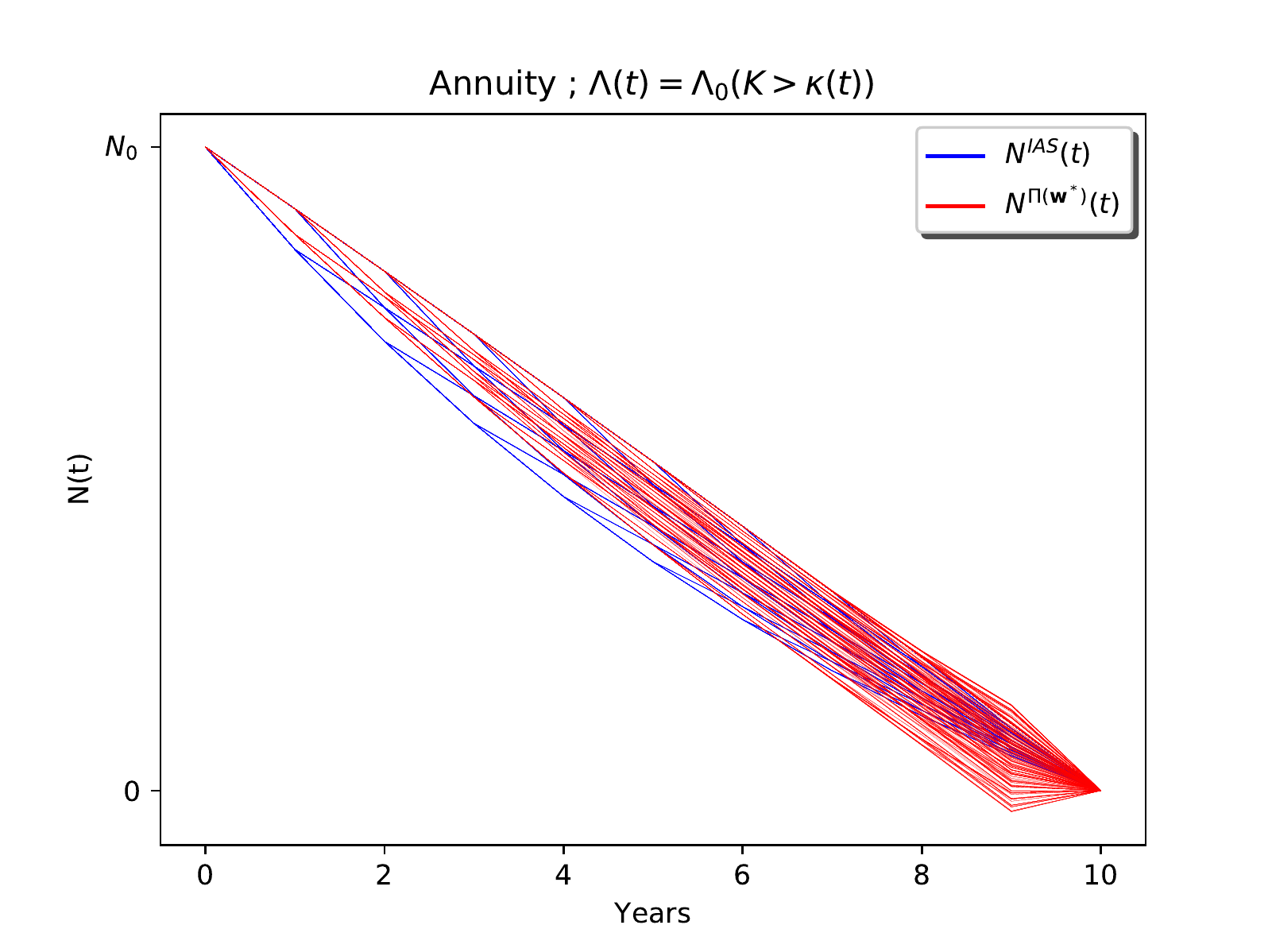} \hfill
    \includegraphics[width=0.495\textwidth]{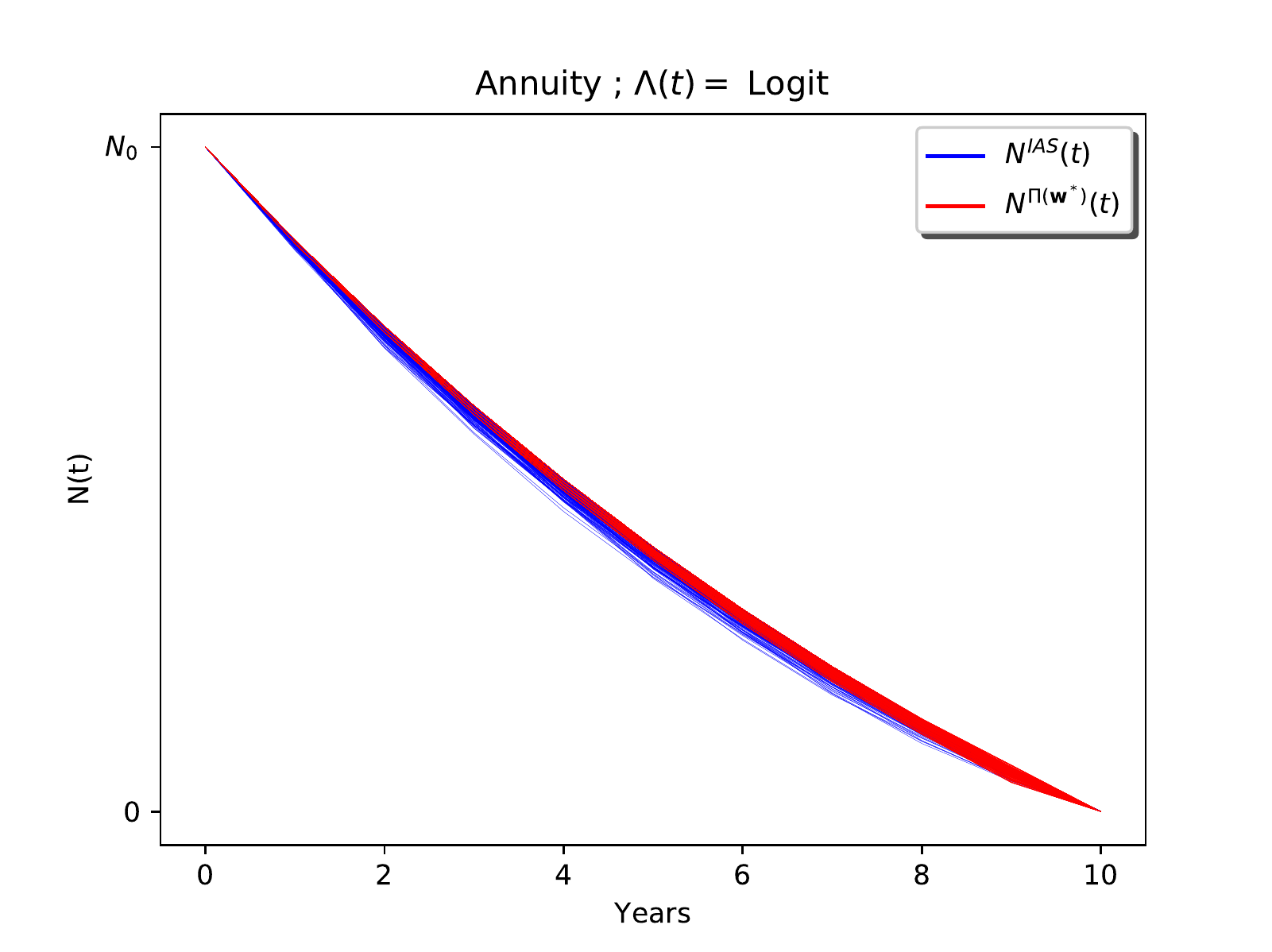}
    \caption{Examples how the calibrated hedging portfolio composed of an amortizing swap and nine swaptions (red) replicates the notional of the IAS (blue). Left: fully rational refinancing incentive. Right: S-shaped function. Top: bullets. Bottom: annuities.}
    \label{fig:HedgingExamplesOnNotionals}
\end{figure}

Table \ref{tbl:CompositionPortfolioNineSwaptions} presents the composition of the portfolio of swaptions for the bullet and annuity mortgages, expressing the corresponding prices in basis points, while Figure \ref{fig:HedgingExample_BulletFewPaths} presents an example of the replication quality of the calibrated portfolio, where the red colored paths of $N_{\Pi(\mathbf{w}^*)}$ resemble the  blue colored paths of $N_{\text{IAS}}$ closely. Figure \ref{fig:HedgingExamplesOnNotionals} illustrates the results of the calibration of the nine swaptions by which a mortgage of ten years can be hedged.

Four aspects, one in each of the plots, will be discussed, starting from the top-left corner. In the case of a bullet with a fully rational prepayment rule, there is a mismatch between $N_{\text{IAS}}$ and $N_{\Pi}$ close to maturity time. Since the prepayment rate is proportional to the outstanding notional, the blue grid shrinks slightly, close to maturity. However, this effect is not reflected in the hedge because the swaptions can not be ``partly'' exercised. In the second graph of the first line, the bullet with an S-shaped function for the RI, the red paths do not touch the bottom-right set of blue paths. Since we  minimize the sum of the squared differences in a path-wise fashion, these paths do not have significant weights. The last two considerations concern the annuity, i.e., the graphs in the second line. In the left-side graph, close to maturity time, an awkward effect is visible, caused by excessive amortization of the swaptions. The swaptions bought to replicate the first five years appear to negatively affect the performance close to maturity time, indicating that negative notionals for the last swaptions would be necessary. However, this effect is mitigated in the practical setting (i.e., in the last graph) because the paths of the annuity using an S-shaped RI curve are close. Thus, we conclude that not too many swaptions are required as the underlying assets' risk becomes ``linear'' in the case of the annuity mortgage.
\begin{table}[h]
\centering
\footnotesize
\begin{tabular}{ccccccccc|c}
\toprule
1Y-9Y & 2Y-8Y & 3Y-7Y & 4Y-6Y & 5Y-5Y & 6Y-4Y & 7Y-3Y & 8Y-2Y & 9Y-1Y & Total \\
(bps) & (bps) & (bps) & (bps) & (bps) & (bps) & (bps) & (bps) & (bps) & (bps)  \\
\midrule
1.16 & 1.34 & 1.41 & 1.43 & 1.40 & 1.34 & 1.23 & 0.99 & 0.61 & 10.95 \\
0.61 & 0.66 & 0.58 & 0.48 & 0.35 & 0.25 & 0.12 & 0.01 & -0.07 & 3.01 \\
\bottomrule
\end{tabular}
\caption{Composition of the calibrated portfolio of swaptions for the bullet and annuity mortgages. The weights multiply the prices of the swaptions and have been divided by the notional of the mortgage to give an insight into the basis points that should be invested to apply the hedging strategy. Top: Bullet. Bottom: Annuity.}
\label{tbl:CompositionPortfolioNineSwaptions}
\end{table}

Since in the bullet mortgage, the outstanding notional does not decrease with time, the impact of prepayments is more pronounced than for annuity mortgages, especially considering the increasing uncertainty as time proceeds. In annuity mortgages, on the other hand, the outstanding notional decreases with time and therefore, the impact of prepayments is reduced. Moreover, since convexity (i.e., nonlinearity) is a function of volatility and time, it is less pronounced in the case of annuity mortgages.

This does not mean that hedging the prepayment risk is not necessary since it is not assured that the non-rational prepayments are always greater than the rational prepayments. The second noteworthy effect is that the annuity contract exhibits smaller deviations in both experiments from the at-the-money price, which is equal to zero. This is due to the fact that annuities have much less contractual freedom (which will be explained below). Therefore, they are much less impacted by an incentive.
\subsection{Price of the prepayment option \label{subsection:PricePrepaymentOption}}
In principle, the prepayment option price is simply the difference of the mortgage structure with and without prepayments. This insight is somewhat misleading, as there is always a minimum level of prepayment. Those prepayments do not constitute a risk, and, therefore, they will be excluded from the pricing of the prepayment option and hedged with linear instruments. The distinction between linear and non-linear risk is useful for hedging purposes, where one can address the risk to the tradeable instruments and the pricing itself. Equation (\ref{eqn:PortfolioDef_LongAS_ShortSwaptions}) shows that the linear and non-linear effects move in different directions, which explains the positive and negative prices in Table \ref{tbl:NumericalResults:StochasticNotional}. In the case of the fully rational prepayment function, a negative price was obtained. Because the mortgage at time $t_0$ was at the money and the linear part of the hedge has no costs, only the swaptions play a role then, leading to the negative price. However, in reality, the upper bound of the notional is much lower than the curve representing the case without prepayment, and the linear hedge is important too. The price of the IAS appears to be positive, because the value of the AS in (\ref{eqn:PortfolioDef_LongAS_ShortSwaptions}) is larger than the price of the swaptions. This observation, however, is strongly dependent on the shape of the yield curve, i.e., one would expect the sign to flip in the case of an inverted yield curve.

\subsection{Comparison of different hedging portfolios \label{subsection:AssessingHedgingPerformances}}
\begin{figure}[h]
    \centering
    \includegraphics[width=0.495\textwidth]{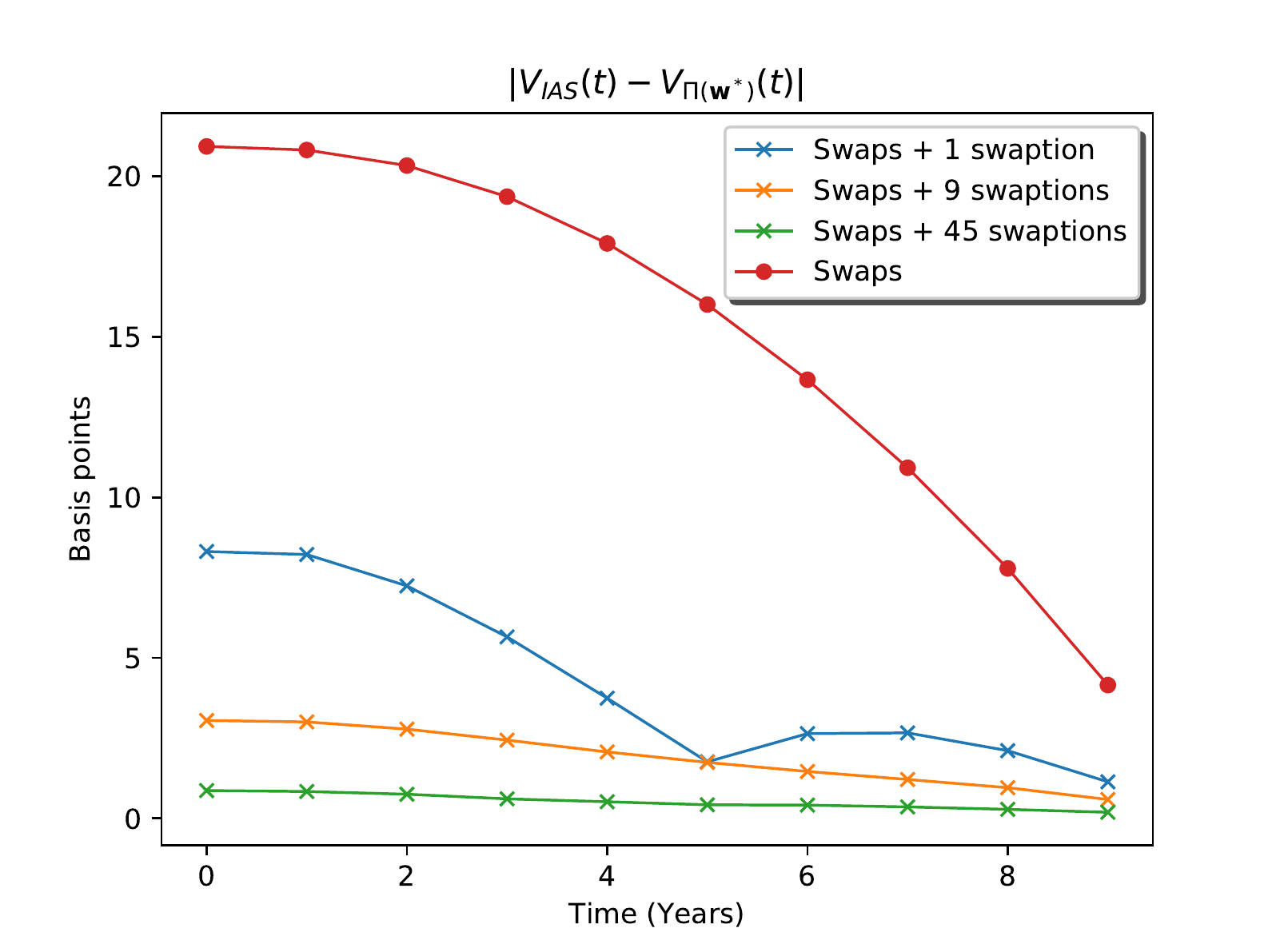} \hfill
    \includegraphics[width=0.495\textwidth]{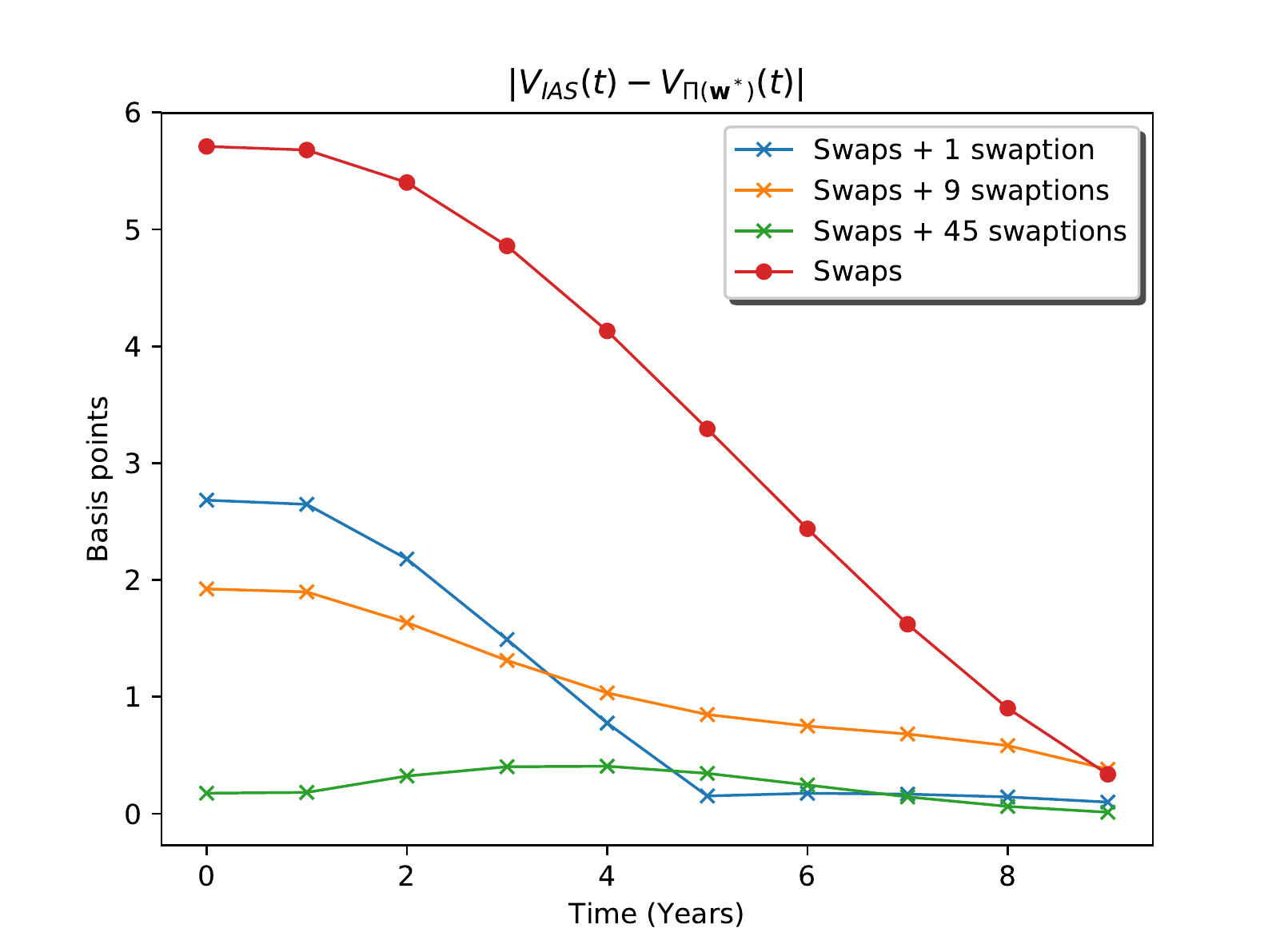}
    \caption{Comparison of the performance of different static hedge strategies. The complete replication employs all swaptions available, while already half of the risk is covered by only one swaption. The nine swaptions on the counter-diagonal appear to be an optimal compromise, allowing for accurate hedging without using an excessive number of instruments. Left: Bullet. Right: Annuity.}
    \label{fig:HedgingComparisonValues}
\end{figure}
We assess the performance of the different hedging strategies to provide an insight into the swaptions that are most important in replicating the IAS. We only have an analytic calibration for the counter-diagonal's swaptions, so a numerical minimization will be performed for the other portfolios. The value of the IAS at time $T_k$ is defined as:
\begin{equation*}
\begin{split}
    V_\text{IAS}(T_k) & = \mathbbm{E}^\mathbbm{Q} \left[ \sum_{i=k+1}^{M} \tau_i \frac{N(T_{i-1};\Lambda(T_{i-1}))}{M(T_i)}   \cdot \big( K - L(T_{i-1};T_{i-1},T_i) \big) \big | \mathcal{F}(T_k) \right] \\[1.0ex]
    \approx & \frac{1}{N_{Sim}} \sum_{j=1}^{N_{Sim}} \sum_{i=k+1}^{M} \tau_i N_{\text{IAS}}^{(j)} \frac{M^{(j)}(T_k)}{M^{(j)}(T_i)}   \cdot \big( K - L^{(j)}(T_{i-1};T_{i-1},T_i) \big).
\end{split}
\end{equation*}
We analyze the accuracy of the hedge by computing the difference,
$
    \left| V_\text{IAS}(T_k) - \Pi(T_k,\mathbf{w^*}) \right|,
$
for each time $T_k$, $k=0,...,M-1$. Figure \ref{fig:HedgingComparisonValues} shows the linear hedge performance as well as the non-linear hedge composed of swaps and a varying number of swaptions. The portfolio with only one swaption, $5Y-5Y$, primarily helps to replicate the 10y mortgage, with its peak performance, as expected, in the fifth year. The portfolio with nine swaptions from the diagonal exhibits the optimal compromise of the full-replication and the linear hedge strategy. In practice,  financial institutions mainly hedge the prepayment risk using a dynamic linear hedge. Nevertheless, our results give insight into the benefits of using a small number of swaptions, not only for a static hedge but also for a dynamic hedge. To clarify this, one should analyze the Greeks of the IAS and of $\Pi(\mathbf{w^*})$.

In theory, one would use all information that the market provides to accurately calibrate model parameters. In practice, this is challenging (if not impossible), especially with the Hull-White or CIR++ short-rate models. A comprehensive calibration is not within reach because only the long-term average is time-dependent in these models, while the speed of mean-reversion and volatility parameters are assumed to be constant. In the case of the CIR++ model, the long-term average is constant, but the shift introduced to recover negative rates effectively acts as a time-dependent average.
When a subset of instruments has been chosen, a natural choice to calibrate a short-rate model are the market quotes of the instruments that the simulation will most often use, meaning selecting an ``area'' of the volatility matrix. We focus on the swaptions at the counter-diagonal (or the diagonal, if we consider Table \ref{tbl:AllPossibleSwaptions}), so those should also be selected to calibrate the interest rate models. This procedure makes sense because the refinancing incentive depends on the swap rates with these maturities and tenors for the mortgage. Nevertheless, this choice is still too restrictive to achieve accurate  results because the mentioned short-rate models cannot model the implied volatility smile/skew that the market exhibits for these swaptions. One way to overcome this issue is to model the volatility as a time-dependent function, which is beyond the scope of this article. Here, we have selected a smaller set of swaptions. For a mortgage of ten years, the calibrated parameters are reported in Table~\ref{tbl:CalibratedParametersHullWhiteCIR}, while the calibrated prices and implied volatilities are shown in Table \ref{tbl:CalibratedSwaptionsHullWhiteCIR}.
\begin{table}[h]
\centering
\footnotesize
\begin{tabular}{cc|cccc}
\toprule
$\lambda^{\text{HW}}$ & $\eta^{\text{HW}}$ & $\lambda^{\text{CIR}}$ & $\eta^{\text{CIR}}$ & $\theta^{\text{CIR}}$ & $x_0$  \\
\midrule
$0.264$ & $0.017$ & $0.185$ & $0.039$ & $0.184$ & $0.079$\\
\bottomrule
\label{table}
\end{tabular}
\caption{Calibration of the Hull-White and CIR++ models.}
\label{tbl:CalibratedParametersHullWhiteCIR}
\end{table}

\begin{table}[h]
\centering
\footnotesize
\begin{tabular}{ccccccc}
\toprule
Swaption & $\sigma_{\text{Market}}$ & $\sigma_{\text{HW}}$ & $\sigma_{\text{CIR}}$ & $V_{\text{Market}}$ & $V_{\text{HW}}$ & $V_{\text{CIR}}$ \\
(Maturity-Tenor) & (bps) & (bps) & (bps) & (bps) & (bps) & (bps) \\
\midrule
1Y-10Y & 46.31 & 53.90 & 48.00 & 177.96 & 207.13 & 184.45 \\
3Y-7Y & 56.28 & 56.22 & 55.40 &  261.61 & 261.34 & 257.53  \\
5Y-5Y & 61.98 & 57.37 & 60.28 & 262.28 & 242.75 & 255.11  \\
7Y-3Y & 64.79 & 61.75 & 66.36 & 191.63 & 182.62 & 196.27  \\
9Y-1Y & 64.89 & 69.79 & 74.10 & 71.27 & 76.66 & 81.39  \\
\bottomrule
\end{tabular}
\caption{Results of the calibrated swaptions.}
\label{tbl:CalibratedSwaptionsHullWhiteCIR}
\end{table}

A consistent choice of instruments to construct the yield curve would imply that the swap rates selected are the underlying instruments of the swaptions. However, it is common practice to show the sensitivities to spot instruments because the ``forward Delta-profile'' appears more challenging  to interpret, so we use spot swap rates instead.
\subsection{Approximation of the Greeks}
\begin{figure}[h]
    \centering
    \includegraphics[width=0.495\textwidth]{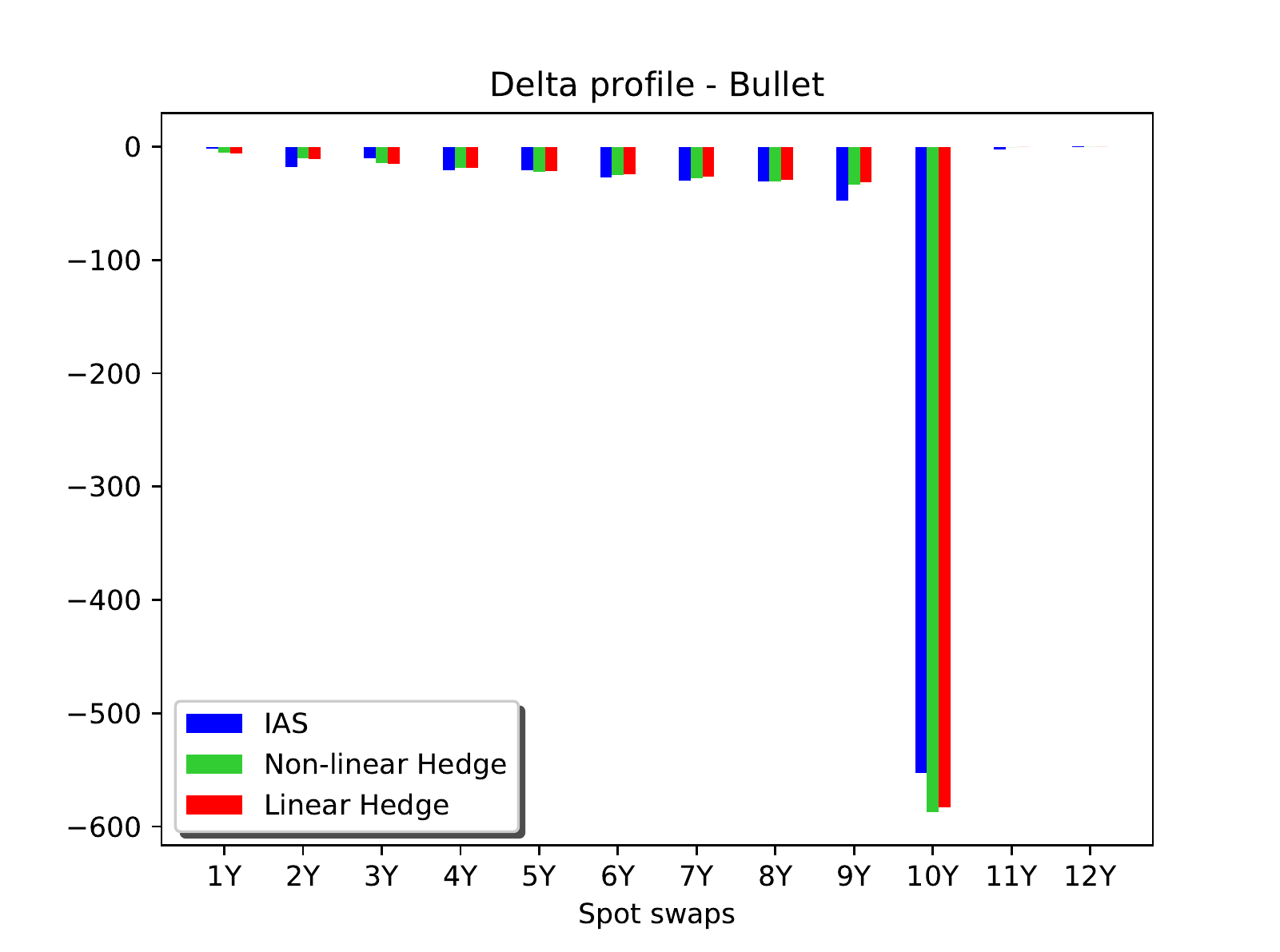} \hfill
    \includegraphics[width=0.495\textwidth]{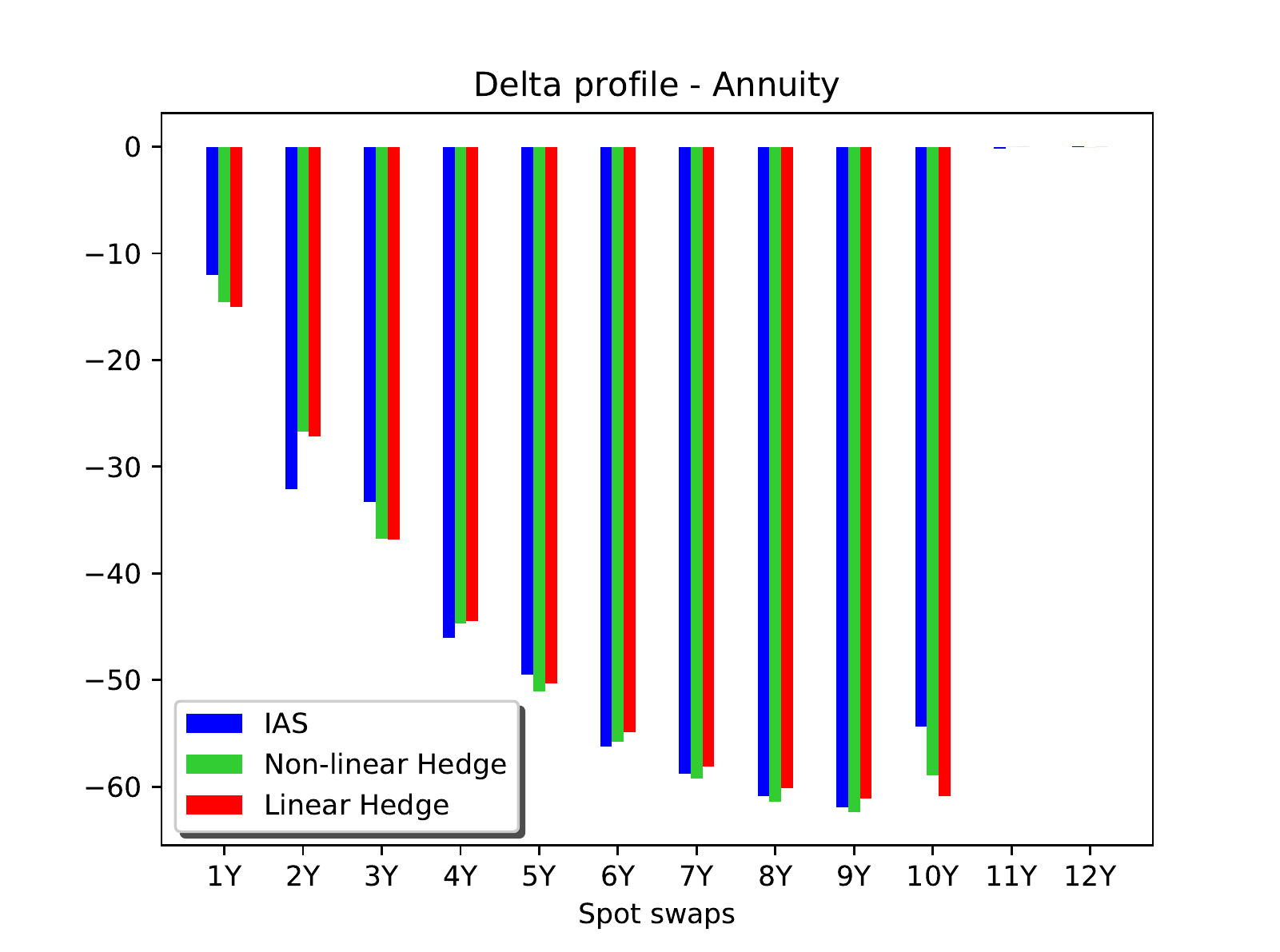}
    \caption{Delta profile of a bullet (left) and an annuity (right). On the x-axis are the spot swap rates used to construct the yield curve. The Delta profile of the IAS is well-approximated by the linear and non-linear hedge (which also has a linear component).  The notional of the mortgage $N_0=1$ million.}
    \label{fig:DeltaProfilesBulletAnnuity}
\end{figure}

We compute the Greeks of the IAS numerically by approximating them with finite differences. The instruments to which we can show the sensitivity are the spine swap rates used to construct the yield curve, which affects $\theta^{\text{HW}}$, and the implied volatilities on which the parameters $\left(\lambda^{\text{HW}}, \eta^{\text{HW}} \right)$ have been calibrated.

Variation of the price of a derivative with respect to these quantities implies that the price of the IAS is a function of them, so
\begin{equation*}
    V_{\text{IAS}}(t_0) = V_{\text{IAS}}(t_0;S_{m,n},\sigma_{i,j}^{\text{Market}}).
\end{equation*}
Delta, the first-order sensitivity, is the Greek that is primarily used to hedge. We approximate it by central differences:
\begin{equation*}
    \Delta_{\text{IAS}}(S_{m,n}) = \frac{\partial V_{\text{IAS}}(t_0;S_{m,n}) }{\partial S_{m,n} } \approx     \frac{V_{\text{IAS}}(t_0;S_{m,n}+h) - V_{\text{IAS}}(t_0;S_{m,n}-h)}{2h}.
\end{equation*}

Gamma measures the rate of change of Delta, and is computed as,
    \begin{equation*}
        \Gamma_{\text{IAS}}(S_{m,n}) = \frac{\partial^2 V_{\text{IAS}}(t_0;S_{m,n}) }{\partial S_{m,n}^2 } \approx \frac{V_{\text{IAS}}(t_0;S_{m,n}+h) - 2 V_{\text{IAS}}(t_0;S_{m,n}) + V_{\text{IAS}}(t_0;S_{m,n}-h)}{h^2}.
    \end{equation*}

Vega measures the sensitivity to the volatility, and we determine Vega by,
    \begin{equation*}
        \mathcal{V}_{\text{IAS}}(\sigma_{i,j}) = \frac{\partial V_{\text{IAS}}(t_0;\sigma_{i,j}) }{\partial \sigma_{i,j} } \approx     \frac{V_{\text{IAS}}(t_0;\sigma_{i,j}+h) - V_{\text{IAS}}(t_0;\sigma_{i,j})}{h}.
    \end{equation*}

\begin{figure}[h]
    \centering
    \includegraphics[width=0.495\textwidth]{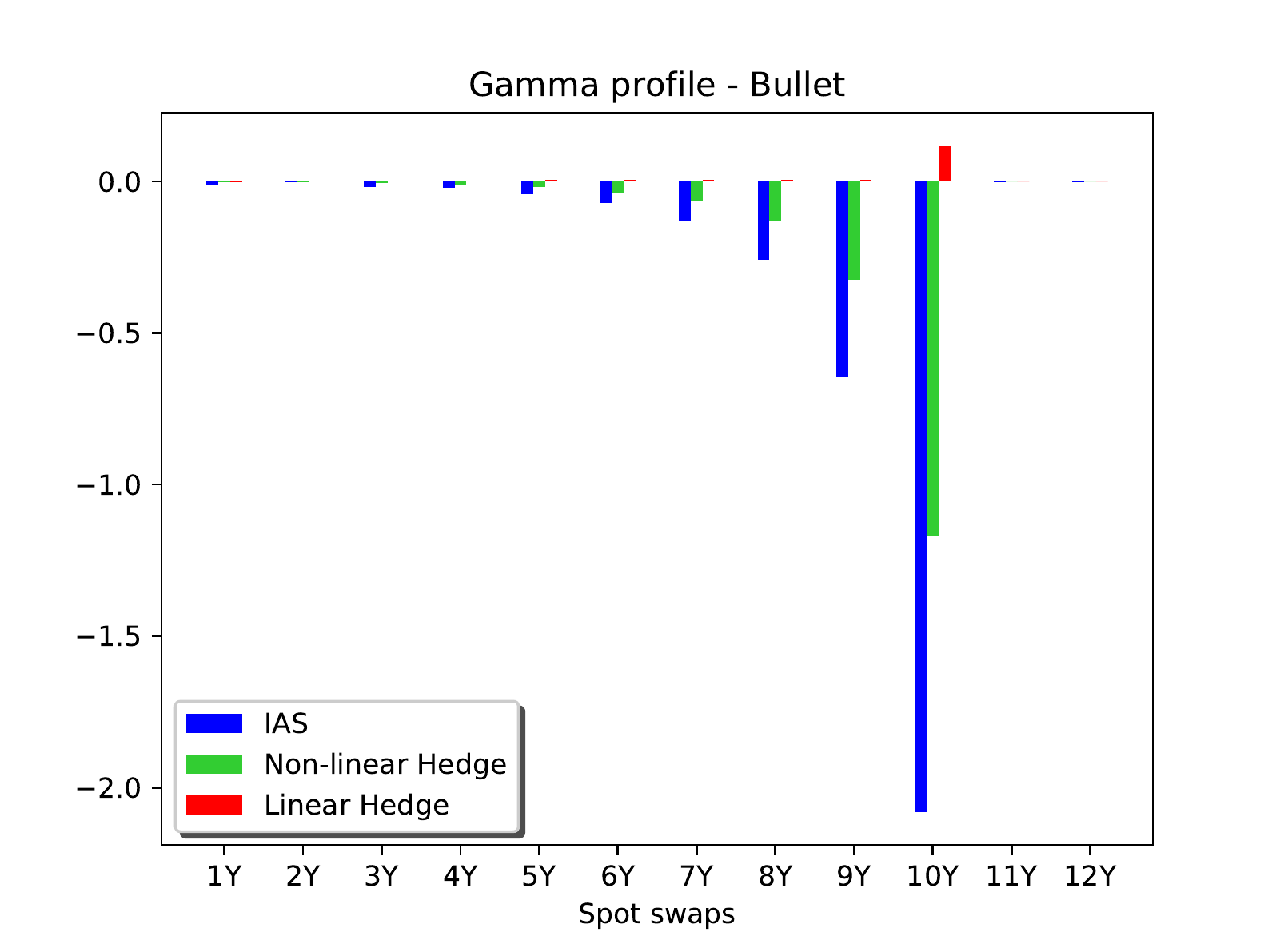} \hfill
    \includegraphics[width=0.495\textwidth]{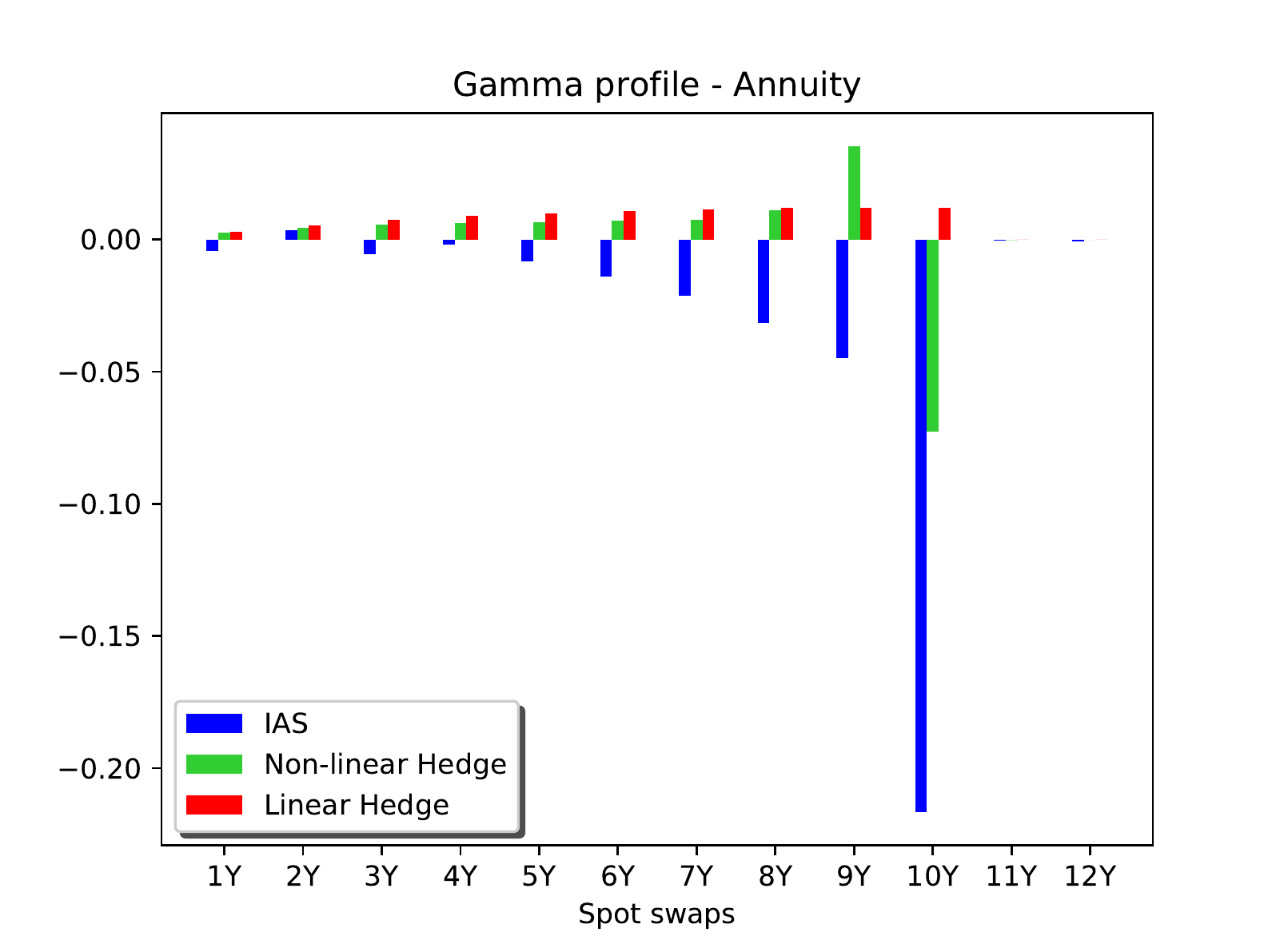}
    \caption{Gamma profile of a bullet (left) and an annuity (right). On the x-axis are the spot swap rates used to construct the yield curve. The Gamma profile of the bullet shows higher values compared to the annuity, and the advantages of the non-linear hedge are clear compared to the linear hedge. The Gamma profile of the annuity is strange, however the values are negligible. The notional of the mortgage $N_0=1$ million.}
    \label{fig:GammaProfilesBulletAnnuity}
\end{figure}

The pricing of the linear and non-linear hedge strategies can be done analytically. The Delta profiles of the bullet and of the annuity appear to be replicated, while the performance on the Gamma profiles is less accurate.  As expected, most of the Delta of a bullet with a maturity of ten years is on the swap rate $S_{0,10}$, because, since a bullet does not involve any repayments, it is closer to a plain-vanilla swap than an annuity is. On the other hand, the Delta of an annuity is spread over the instruments because its amortization plan is based on each of them. For the annuity, the Gamma of the hedge seems to be different from the Gamma of the IAS. However, notice that the magnitude of the Gamma embedded in an annuity is small, especially when compared to the bullet. This behaviour was expected when looking at the paths of the notional of an annuity (for example, Figure \ref{fig:HedgingExamplesOnNotionals}); the cloud of paths is not as large as the one of the bullet. Thus there is no need to include many swaptions. Finally, the Vega profiles are partially replicated for the bullet and the annuity by the non-linear hedge, while, of course, the swaps did not return any Vega value, emphasizing the necessity to include the non-linear hedge in the protection against the prepayment risk.

\begin{figure}[h]
    \centering
    \includegraphics[width=0.495\textwidth]{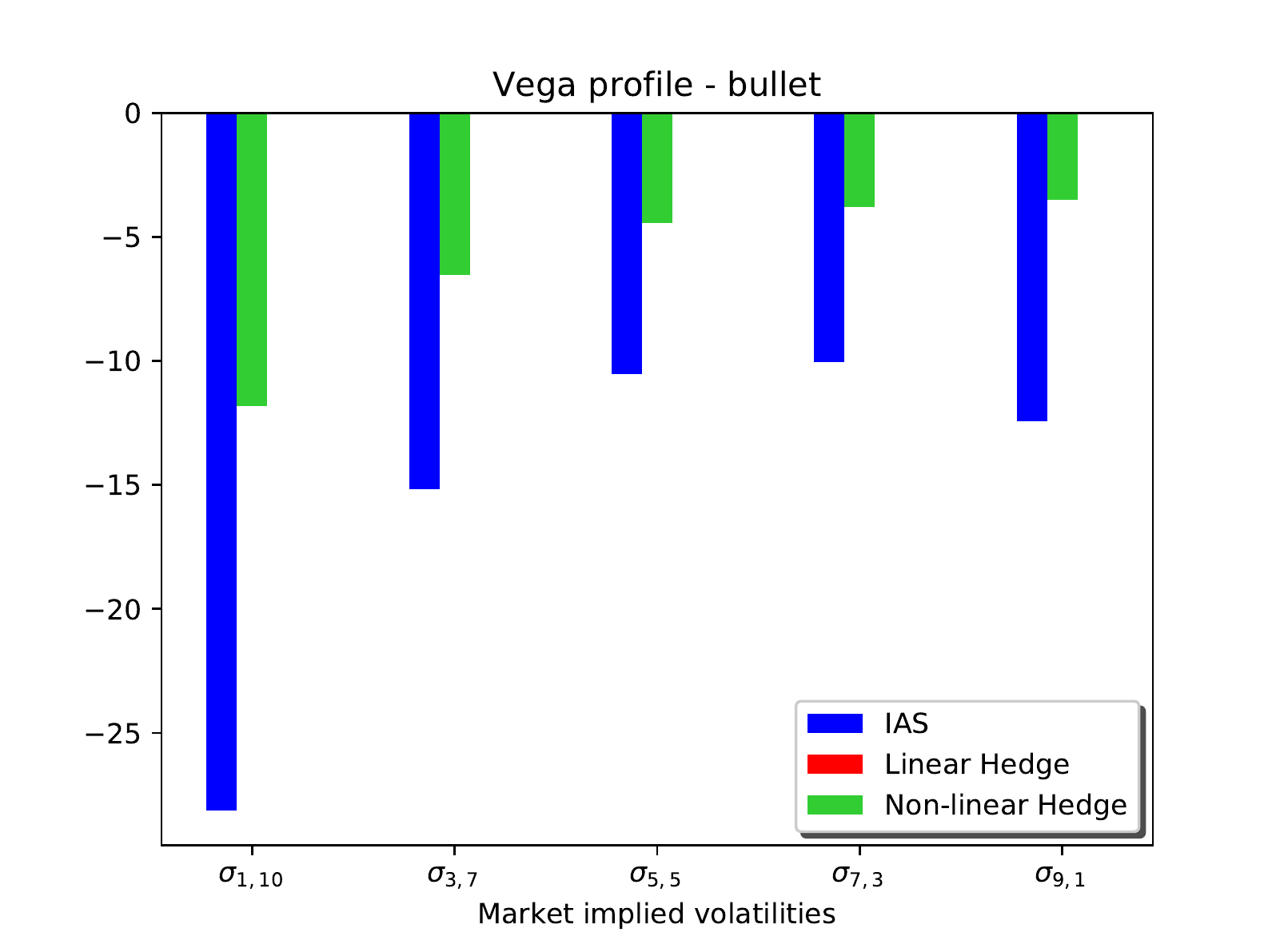} \hfill
    \includegraphics[width=0.495\textwidth]{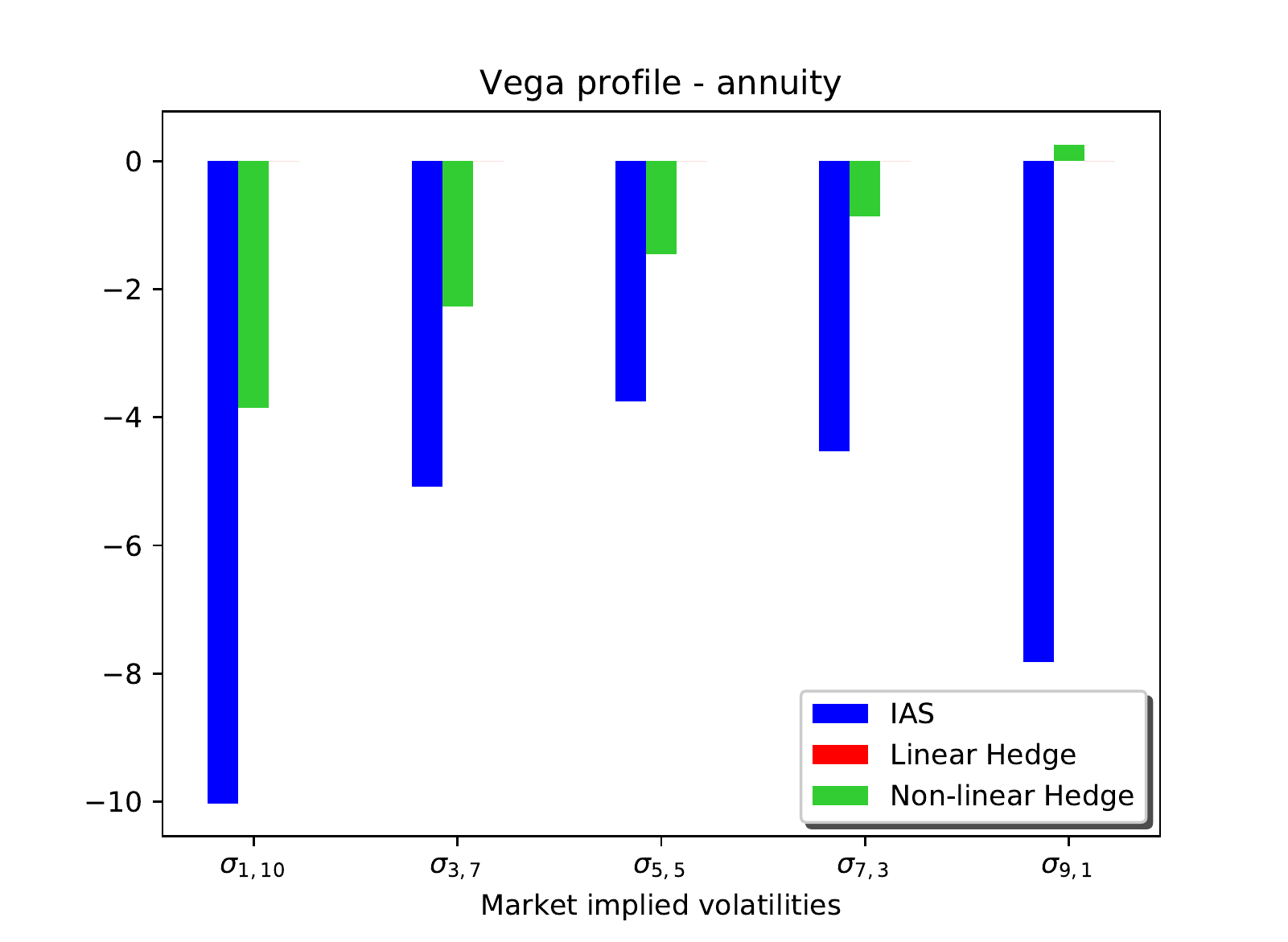}
    \caption{Vega profile of a bullet (left) and an annuity (right), with on the x-axis the implied volatilities used to calibrate the Hull-White model. As expected, the non-linear hedge approximates the Vega profile of the IAS, while the linear hedge does not show any sensitivity to changes in the market volatility. Similar to  Gamma,  the bullet is more sensitive than the annuity here. The notional of the mortgage $N_0=1$ million.}
    \label{fig:VegaProfilesBulletAnnuity}
\end{figure}

\subsection{Calibrating the hedging portfolio on Gamma}
\begin{figure}[h]
    \centering
    \includegraphics[width=0.495\textwidth]{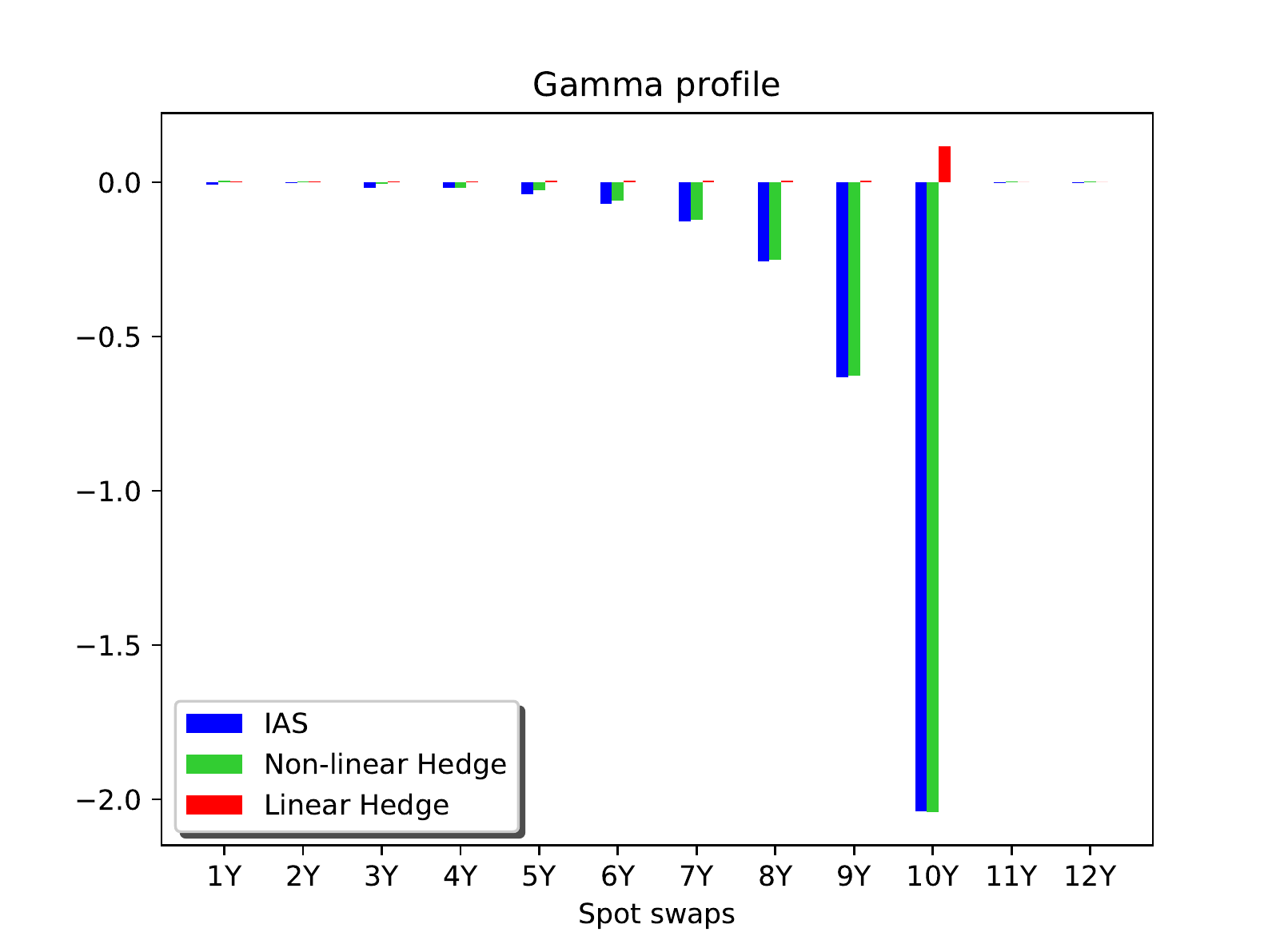} \hfill
    \includegraphics[width=0.495\textwidth]{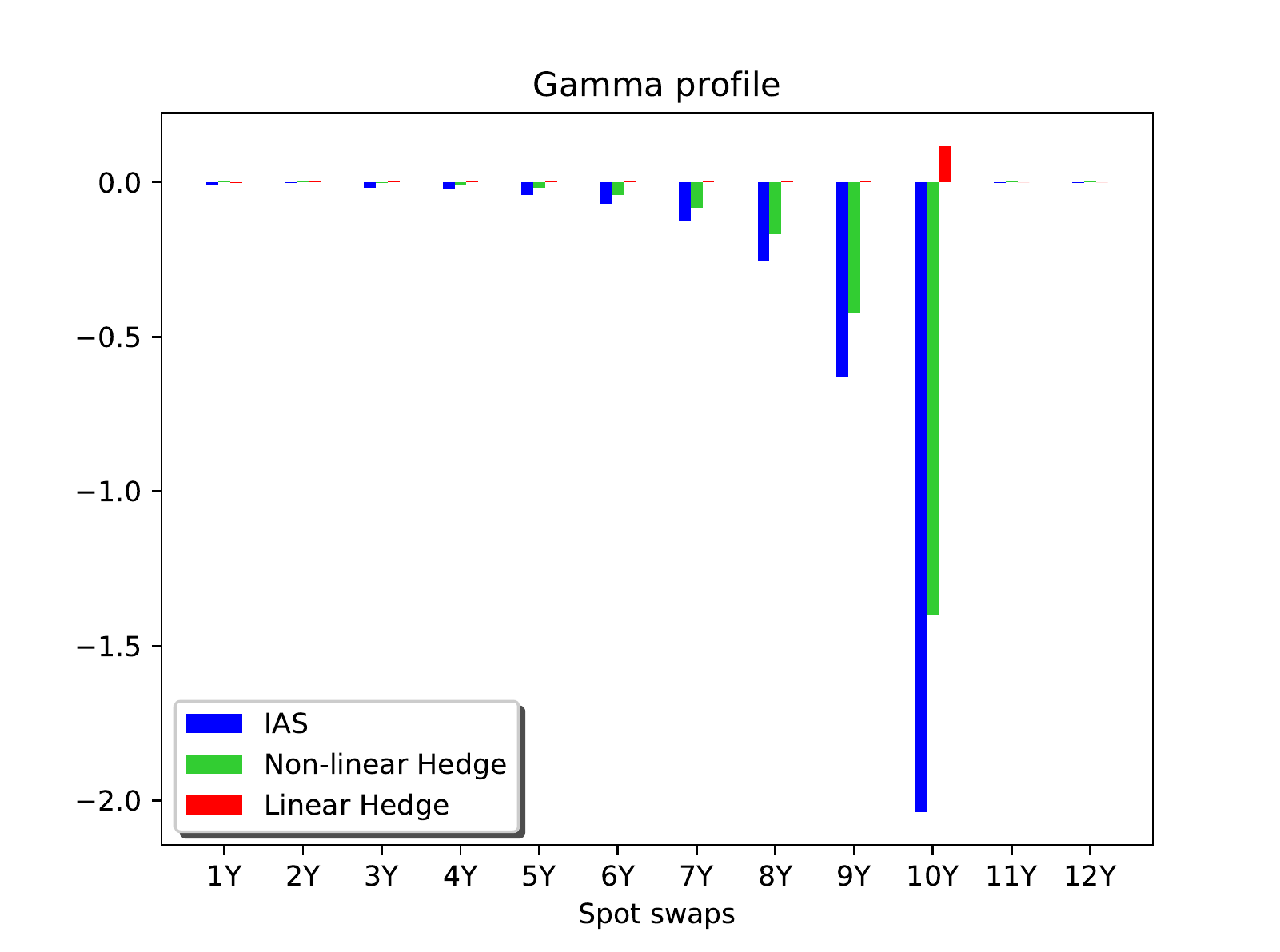} \\
    \includegraphics[width=0.495\textwidth]{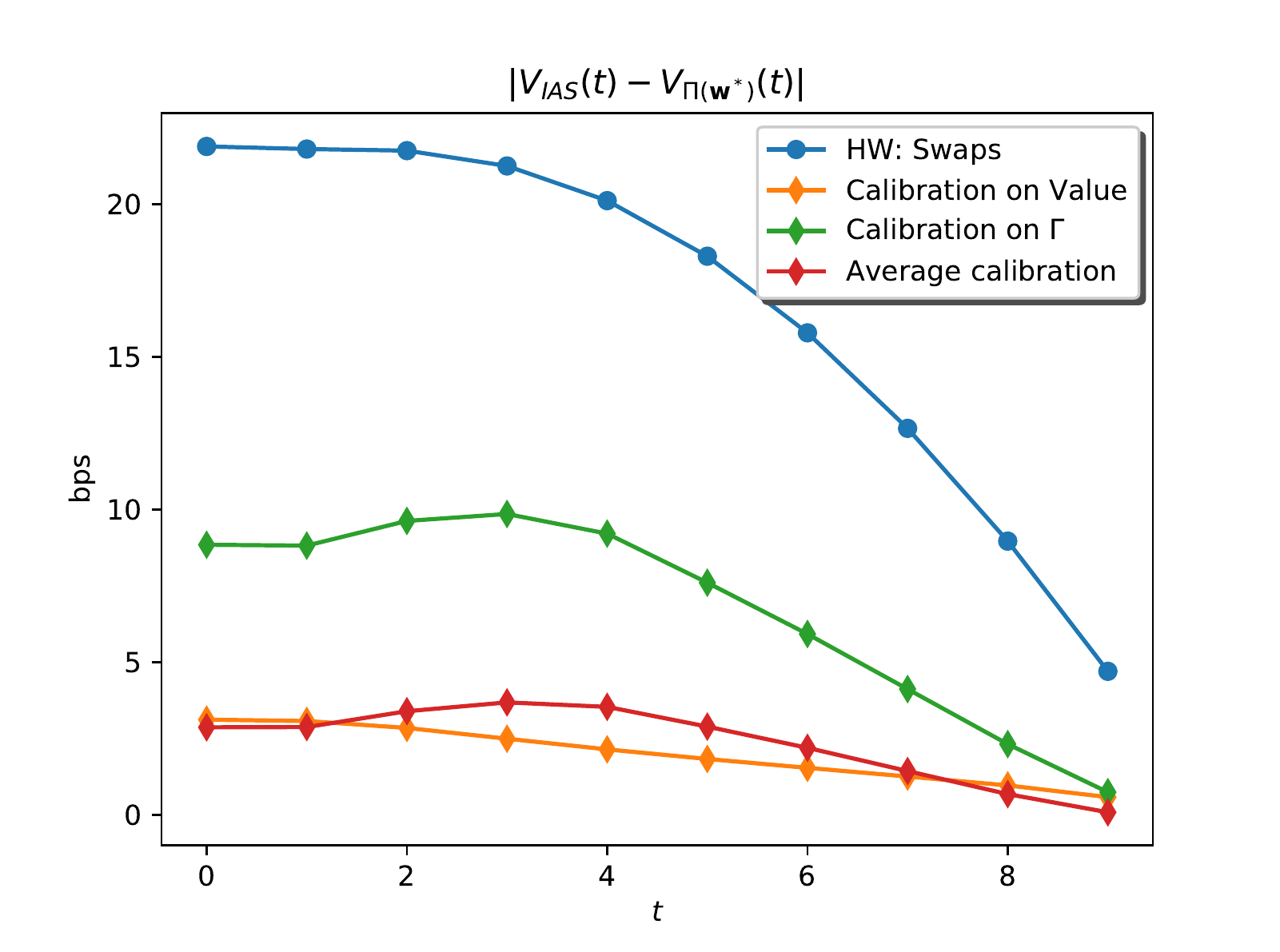}
    \caption{Top left: Gamma profiles using $\Pi(\mathbf{w}_\Gamma^*)$. Top right: Gamma profiles using $\Pi(\widetilde{\mathbf{w}})$. Bottom: Comparison of the approximation of the price of $\Pi(\mathbf{w}^*)$ (calibration on value), $\Pi(\mathbf{w}_\Gamma^*)$ (calibration on Gamma) and $\Pi(\widetilde{\mathbf{w}})$ (average calibration).}
    \label{fig:ComparisonDifferentCalibrationOnGamma}
\end{figure}

Since the portfolio of nine swaptions offsets the Gamma of the IAS only partly, we analyze how the composition of the portfolio changes when the calibration is based on the replication of the Gamma of the IAS. Changing the linear part of the hedge would be useless for the offset of Gamma. We therefore calibrate the swaptions obtaining the weights from (\ref{eqn:MinimizationFunctionWeightsSwaptionGeneral}) using the functional,
\begin{equation}
    F(\textbf{w}) = \left\lVert \Gamma_{\text{IAS}} - \Gamma_{\Pi(\mathbf{w})} \right\lVert_2,
\label{eqn:MinimizationFunction_Gamma}
\end{equation}
where $\Gamma_{\text{IAS}}$ and $\Gamma_{\Pi(\mathbf{w})}$ are vectors with as entries the Gamma values of the IAS and of the non-linear hedge. The minimization returns a vector of weights, $\mathbf{w}_\Gamma^*$, which determines a new composition of the portfolio, as shown in Table \ref{tbl:CompositionPortfolioNineSwaptions_CalibrationGamma}. Notice that this solution costs almost twice the amount of the original $\mathbf{w}^*$, which must result in a worse replication of the price of the IAS. Thus, we find a compromise in a naive way, by ``averaging'' the weights of the calibrations on the prices ($\mathbf{w}^*$) and on Gamma ($\mathbf{w}_\Gamma^*$), by using,
\begin{equation*}
    \widetilde{\mathbf{w}} = \frac{\mathbf{w}^* + \mathbf{w}_\Gamma^*}{2}.
\end{equation*}
As Figure \ref{fig:ComparisonDifferentCalibrationOnGamma} shows, the full replication of the Gamma leads to a mismatch in the prices, however, the average portfolio $\widetilde{\mathbf{w}}$ keeps the error on the price below five basis points and provides a better replication of the Gamma profile of the IAS.

\begin{table}[h]
\centering
\footnotesize
\begin{tabular}{ccccccccc|c}
\toprule
1Y-9Y & 2Y-8Y & 3Y-7Y & 4Y-6Y & 5Y-5Y & 6Y-4Y & 7Y-3Y & 8Y-2Y & 9Y-1Y & Total \\
(bps) & (bps) & (bps) & (bps) & (bps) & (bps) & (bps) & (bps) & (bps) & (bps)  \\
\midrule
1.16 & 1.34 & 1.41 & 1.43 & 1.40 & 1.34 & 1.23 & 0.99 & 0.61 & 10.95 \\
1.76 & 0.00 & 3.36 & 2.65 & 3.53 & 3.51 & 3.27 & 2.85 & 1.80 & 22.78 \\
\bottomrule
\end{tabular}
\caption{Composition of the calibrated portfolio of swaptions for a bullet with two different calibrations. The weights multiply the prices of the swaptions and have been divided by the notional of the mortgage to show the basis points that should be invested to apply the hedging strategy. Top: calibration performed by minimizing (\ref{eqn:MinimizationFunction_SumTimeAverageSimulations}) that returned $\mathbf{w}^*$. Bottom: calibration performed by minimizing (\ref{eqn:MinimizationFunction_Gamma}) that returned $\mathbf{w}_\Gamma^*$.}
\label{tbl:CompositionPortfolioNineSwaptions_CalibrationGamma}
\end{table}

\section{Conclusions}\label{6}
We have investigated methods to price and hedge a portfolio of mortgages, focusing on the prepayment risk. After introducing the prepayment option and different mortgage contracts, we explained the importance of predicting the prepayment rate by defining a link between the refinancing incentive and the financial instruments in the market. The resulting framework enables us to model a stochastic environment, where the paths of an interest rate model will define the paths of the notional of a mortgage. The advantages of this methodology are twofold. First of all, a risk-neutral evaluation appears viable. Secondly, due to the volatility implied by the non-linear instruments
used to calibrate the interest rate model we extended the hedging strategy with non-linear instruments. Currently, it is common to hedge prepayment risk with only linear instruments. However, this is not in line with the nature of the prepayment option, which, being an option, gives rise to non-linear risk. By implementing the pricing model and calibrating it to actual market data, one can test the different hedge portfolios on the simulated market states.
\small
\bibliographystyle{abbrv}
\bibliography{report}

\appendix
\section{Market Data}
The market volatilities $\sigma_{m,n}^{\text{N}}$ used to calibrate the interest rate models are in Table~\ref{tbl:SwaptionVolatilityMatrix}. In  the columns, we find the quotes for swaptions with different maturities but with the same duration for the underlying swaps. We read the quotes for the swaptions with a fixed maturity and an increased length of the underlying swap in rows. The payments for the fixed and floating lengths are supposed to occur with the same frequency every semester.
\begin{table}[h]
\centering
\footnotesize
\begin{tabular}{c|cccccccccccc}
\toprule
     & 1Yr   & 2Yr   & 3Yr   &   4Yr &   5Yr &   7Yr &  10Yr & 12Yr  & 15Yr  & 20Yr  & 25Yr  & 30Yr \\
\midrule
1Mo  & 8.86  & 14.61 & 20.09 & 25.55 & 29.49 & 32.30 & 33.58 & 34.88 & 36.28 & 37.92 & 39.46 & 40.86 \\
3Mo  & 11.05 & 16.25 & 21.74 & 27.52 & 31.34 & 34.65 & 36.71 & 38.00 & 39.30 & 40.60 & 41.54 & 42.25 \\
6Mo  & 14.75 & 20.81 & 26.38 & 31.74 & 35.37 & 38.63 & 40.71 & 41.84 & 43.01 & 43.99 & 44.51 & 44.90  \\
9Mo  & 18.37 & 24.96 & 30.00 & 34.98 & 38.46 & 41.59 & 43.86 & 44.82 & 45.79 & 46.63 & 47.07 & 47.48 \\
1Yr  & 22.40 & 29.34 & 34.17 & 38.25 & 41.26 & 44.01 & 46.31 & 47.15 & 47.97 & 48.68 & 49.07 & 49.39 \\
2Yr  & 37.99 & 42.76 & 46.29 & 48.58 & 49.53 & 51.14 & 52.78 & 53.08 & 53.21 & 53.52 & 53.41 & 53.29 \\
3Yr  & 50.12 & 52.65 & 53.99 & 55.17 & 55.53 & 56.28 & 57.18 & 56.98 & 56.22 & 56.18 & 55.60 & 55.35 \\
4Yr  & 57.30 & 58.56 & 58.98 & 59.35 & 59.27 & 59.80 & 60.07 & 59.31 & 57.96 & 57.42 & 56.43 & 56.04 \\
5Yr  & 61.34 & 61.93 & 62.02 & 62.17 & 61.98 & 62.30 & 62.13 & 60.91 & 59.14 & 58.14 & 56.82 & 56.17 \\
6Yr  & 63.21 & 63.79 & 63.69 & 63.63 & 63.24 & 63.24 & 62.98 & 61.53 & 59.30 & 57.89 & 56.46 & 55.77 \\
7Yr  & 64.41 & 64.95 & 64.79 & 64.72 & 64.26 & 64.13 & 63.56 & 61.99 & 59.49 & 57.76 & 56.15 & 55.36 \\
8Yr  & 64.88 & 65.46 & 65.24 & 64.98 & 64.91 & 64.37 & 63.66 & 62.03 & 59.34 & 57.38 & 55.68 & 54.73 \\
9Yr  & 64.89 & 65.62 & 65.37 & 64.99 & 64.68 & 64.38 & 63.61 & 61.86 & 59.17 & 57.05 & 55.26 & 54.21 \\
10Yr & 64.75 & 65.46 & 65.20 & 64.81 & 64.60 & 64.18 & 63.49 & 61.62 & 58.91 & 56.54 & 54.64 & 53.58 \\
12Yr & 63.62 & 64.43 & 64.33 & 63.71 & 63.34 & 63.01 & 62.17 & 60.37 & 57.63 & 55.04 & 53.09 & 51.90 \\
15Yr & 61.33 & 62.05 & 61.77 & 61.58 & 61.45 & 60.84 & 60.09 & 58.26 & 55.59 & 52.70 & 50.70 & 49.30 \\
\bottomrule
\end{tabular}
\caption{Swaption matrix obtained from Bloomberg on the 23rd of January 2018. Index tenor: 3 months Euribor. Values expressed in basis points for volatilities of ATM swaption, assuming a Normal distribution of the underlying.}
\label{tbl:SwaptionVolatilityMatrix}
\end{table}
\end{document}